\pdfoutput=1  

\documentclass[aps,superscriptaddress,floatfix,nofootinbib,preprintnumbers,eqsecnum,twocolumn]{revtex4-1}
\usepackage[utf8]{inputenc}
\usepackage{lipsum, graphicx}
\usepackage{amssymb,amsmath,multirow,mathtools}
\usepackage{comment,color,placeins,amsmath,amsfonts,enumitem}
\usepackage{cancel}

\usepackage{hyperref}
\usepackage[usenames,dvipsnames]{xcolor}
\hypersetup
{
  colorlinks,
  linkcolor={blue!80!black},
  citecolor={blue!70!black},
  urlcolor={blue!70!black}
}

\makeatletter
\renewcommand*\env@matrix[1][\arraystretch]{%
  \edef\arraystretch{#1}%
  \hskip -\arraycolsep
  \let\@ifnextchar\new@ifnextchar
  \array{*\c@MaxMatrixCols c}}
\makeatother

\def\beq{\begin{equation}}
\def\eeq{\end{equation}}

\newcommand{\il}[1]{\mbox{$#1$}} 

\newcommand{\nn}{\nonumber}
\newcommand{\abs}[1]{\left| #1 \right|}

\newcommand{\p}{\partial}
\newcommand{\pd}[2]{\frac{\partial #1}{\partial #2}} 
\newcommand{\Jsc}{\text{sc}}

\newcommand{\sech}{\text{sech}}
\newcommand{\csch}{\text{csch}}
\newcommand{\thetain}{\theta_{\text{in}}}
\newcommand{\phiin}{\phi_{\text{in}}}

\def\ie{{\it i.e.}\/}
\def\eg{{\it e.g.}\/}
\def\etc{{\it etc}.\/}

\newcommand{\eV}{\text{eV}}

\newcommand{\GeV}{\text{GeV}}

\newcommand{\s}{\text{s}}

\newcommand{\Mpc}{\text{Mpc}}

\newcommand{\Bt}{\widetilde{B}}
\newcommand{\Wt}{\widetilde{W}}
\newcommand{\Ft}{\widetilde{F}}
\newcommand{\Veff}{V_{\text{eff}}}
\newcommand{\feff}{f_{\text{eff}}}
\newcommand{\mphi}{m_{\phi}}
\newcommand{\wphi}{w_{\phi}}

\newcommand{\rhophi}{\rho_{\phi}}
\newcommand{\phidot}{\dot{\phi}}

\newcommand{\dphi}{\delta\phi}
\newcommand{\dphidot}{\delta\dot{\phi}}

\newcommand{\dPphi}{\delta P_{\phi}}
\newcommand{\TD}{T_{\text{D}}}
\newcommand{\TRH}{T_{\text{RH}}}

\newcommand{\LV}{L\hskip -0.15cm\slash}
\newcommand{\GammaLV}{\Gamma_{L\hskip -0.15cm\slash}}
\newcommand{\gammaLV}{\gamma_{L\hskip -0.15cm\slash}}

\newcommand{\BmL}{{B\hspace{-0.03cm}-\hspace{-0.03cm}L}}
\newcommand{\ilBmL}{(B-L)}
\newcommand{\nb}{n_{\phantom{\overline{b}}\hspace{-1.25mm}b}\hspace{-0.5mm}}
\newcommand{\nbbar}{n_{\overline{b}}}
\newcommand{\muBmL}{\mu_\BmL}
\newcommand{\gammaws}{\gamma_{\text{ws}}}
\newcommand{\gammass}{\gamma_{\text{ss}}}
\newcommand{\muws}{\mu_{\text{ws}}}
\newcommand{\Higgs}{\mathcal{H}}
\newcommand{\muH}{\mu_{\mathcal{H}}}
\newcommand{\OmegaCDM}{\Omega_{\text{CDM}}}
\newcommand{\Omegaphi}{\Omega_{\phi}}
\newcommand{\UI}{\text{U}(1)}
\newcommand{\SUII}{\text{SU}(2)}
\newcommand{\SUIII}{\text{SU}(3)}

\graphicspath{{./figs/}} 

\begin{document}
\title{Deformation of Axion Potentials: Implications for Spontaneous Baryogenesis,\\ 
Dark Matter, and Isocurvature Perturbations}
\author{Kyu Jung Bae}
\email{kyujungbae@ibs.re.kr}
\affiliation{\mbox{Center for Theoretical Physics of the Universe, Institute for Basic Science, Daejeon 34126 Korea}}
\author{Jeff Kost}
\email{jeffkost@ibs.re.kr}
\affiliation{\mbox{Center for Theoretical Physics of the Universe, Institute for Basic Science, Daejeon 34126 Korea}}
\author{Chang Sub Shin}
\email{csshin@ibs.re.kr}
\affiliation{\mbox{Center for Theoretical Physics of the Universe, Institute for Basic Science, Daejeon 34126 Korea}}

\preprint{CTPU-PTC-18-40}

\begin{abstract}
We show that both the baryon asymmetry of the universe and dark matter (DM) 
can be accounted for by the dynamics of a single axion-like field.
In this scenario, the observed baryon asymmetry is produced through spontaneous 
baryogenesis---driven by the early evolution of the axion---while its 
late-time coherent oscillations explain the observed DM abundance.
Typically, spontaneous baryogenesis via axions is only successful in 
regions of parameter space where the axion is relatively heavy, rendering 
it highly unstable and unfit as a dark matter candidate.  However, we show that a 
field-dependent wavefunction renormalization can arise which effectively ``deforms''
the axion potential, allowing for efficient generation of 
baryon asymmetry while maintaining a light and stable axion.  
Meanwhile, such deformations of the potential induce non-trivial axion 
dynamics, including a tracking behavior during its intermediate phase of 
evolution.  This attractor-like dynamics dramatically reduces the sensitivity 
of the axion relic abundance to initial conditions and naturally suppresses 
DM isocurvature perturbations.  Finally, we construct an 
explicit model realization, using a continuum-clockwork axion,
and survey the details of its phenomenological viability.
\end{abstract}

\maketitle

\FloatBarrier
\section{Introduction\label{sec:Introduction}}


A wide array of cosmological observations indicate that the universe has a
significant matter-antimatter asymmetry, 
as quantified by the baryon-to-photon ratio~\cite{Iocco:2008va,Ade:2015xua}
\beq\label{eq:etaobs}
\eta_B \equiv \frac{\nb-\nbbar}{n_{\gamma}} = (6.10\pm 0.14)\times 10^{-10} \ ,
\eeq
in which \il{n_B=\nb-\nbbar}
is the baryon-number density and $n_{\gamma}$ is the photon number density.
An essential task of fundamental physics is to explain this figure 
in terms of microphysical processes in the early universe.
Along these lines, a set of necessary conditions 
for the production of baryon asymmetry were obtained 
in a seminal paper by Sakharov~\cite{Sakharov:1967dj}:
(i) violation of baryon number ($B$) 
symmetry\footnote{In circumstances where sphaleron 
processes are in equilibrium, $B$ violation is replaced 
by $\ilBmL$-violation (where $L$ is lepton number), 
so that baryogenesis can also occur via leptogenesis.},
(ii) violation of the discrete $C$ and $CP$ symmetries, 
and (iii) departure from thermal equilibrium.
Typically, satisfying these conditions is a starting point 
for building any model of baryogenesis.
However, it is important to note that the last condition includes
an implicit assumption of $CPT$ invariance.  Indeed, at thermal equilibrium, 
$CPT$ symmetry guarantees that the energy spectra and 
thermal distributions of baryons and antibaryons are equal, thereby 
enforcing \il{\nb=\nbbar}.

By contrast, dynamical scenarios can arise in which $CPT$ is violated 
\emph{spontaneously}, effectively lifting this degeneracy~\cite{Cohen:1987vi,Cohen:1988kt}.  
In this way, baryon asymmetry could be generated at equilibrium, provided 
\mbox{$B$-violating} processes occur at a sufficient rate in the plasma [\ie, that
condition (i) is satisfied].  A model of such ``spontaneous baryogenesis'' 
is typically realized by coupling 
the baryon current $J^{\mu}_B$ to a tensor field which attains 
some non-zero vacuum expectation value (VEV).  
A straightforward example comes in the form of 
a scalar field $\phi$ coupled 
derivatively to the baryon current:
\beq\label{eq:spont_term}
\mathcal{L}_{\text{eff}} \supset \frac{1}{M}\p_{\mu}\phi \cdot J_B^{\mu}  \ ,
\eeq
where $M$ is a cutoff scale.  
It is often reasonable to assume negligible spatial variation in $\phi$,
such that the interaction reduces to \il{M^{-1}\p_0\phi\cdot n_B}.   
In the absence of any scalar field motion, this term has no effect. 
However, as soon as \il{\p_0\phi \neq 0} an energy gap is induced 
between baryon-antibaryon pairs.  In other words, the ``velocity'' 
of the scalar field acts as an \emph{effective chemical potential} for baryon number.
The production of $n_B$ through this mechanism proceeds 
as long as \mbox{$B$-violating} processes are coupled to the thermal bath. 
However, as the universe expands the bath cools, and these eventually decouple,
fixing the baryon asymmetry of the universe (BAU).

A candidate for the scalar $\phi$ can emerge in a variety of contexts: \eg, 
inflatons~\cite{Brandenberger:2003kc,Takahashi:2015ula},
flat directions~\cite{Chiba:2003vp,Takahashi:2003db}, 
radions~\cite{Alberghi:2003ws},
quintessence fields~\cite{Li:2001st,DeFelice:2002ir,Li:2002wd},
scalar curvature~\cite{Davoudiasl:2004gf},
Higgs fields~\cite{Cohen:1991iu,Kusenko:2014lra}, \etc\,
However, scalars with an approximate shift symmetry such as 
pseudo--Nambu Goldstone bosons (pNGB) are
particularly well-motivated in this context~\cite{Carroll:2005dj,Kusenko:2014uta,DeSimone:2016ofp,DeSimone:2016bok}.
These fields may couple linearly to total derivatives, such as 
topological Chern-Simons interactions with SM gauge 
fields \il{\phi\hspace{0.3mm}F\Ft}.  An axion coupling to the weak 
gauge bosons in this way is equivalent 
to Eq.~\eqref{eq:spont_term} due to the electroweak anomaly,
and thus naturally leads to spontaneous baryogenesis.  

In this respect, an axion-like particle~\cite{
    Peccei:1977hh,
    Peccei:1977ur,
    Weinberg:1977ma,
    Wilczek:1977pj,
    Svrcek:2006yi,
    Jaeckel:2010ni}
--- which we shall refer to simply as an ``axion'' --- 
is an attractive candidate for spontaneous baryogenesis models.  
Moreover, it is interesting to consider whether the late-time 
coherent oscillations of the axion field could also play 
the role of dark matter (DM).  
Recent studies~\cite{Kusenko:2014uta,DeSimone:2016ofp} suggest that
axion masses exceeding \il{\mphi\gtrsim 10^5\,\GeV} are necessary to generate
the observed BAU, which would ruin such a prospect.  In particular, a smaller
curvature is associated with the potential of a lighter axion.
This property generally yields a weaker chemical potential $M^{-1}\p_0\phi$
and dynamics triggered at lower temperatures, 
both of which impair spontaneous baryogenesis.
Other proposals have attempted to revive the idea,
such as driving early axion dynamics with 
the Gauss-Bonnet term~\cite{DeSimone:2016bok}, effectively
adding a linear term to the axion potential at early times.
While such a scheme is interesting in that it can be implemented
with the QCD axion, it requires a fine-tuning 
of the misalignment angle, or hierarchical mass scales, 
to obtain the observed DM abundance and prevent 
significant baryonic backreaction or isocurvature perturbations.

In this paper, we describe a novel approach to accommodate both the observed 
baryon asymmetry and DM abundance.  Namely, we consider scenarios in which
``deformations'' to the sinusoidal axion potential arise from a
\emph{field-dependent wavefunction renormalization} $Z(\phi)$.
A variation in $Z(\phi)$ between different regions of the potential 
establishes a mismatch in curvature between those respective regions.
This can have a dramatic effect on axion dynamics and the overall
evolution of the baryon asymmetry and DM abundance. 
In particular, we motivate scenarios in which \il{Z \gg 1} 
toward the minimum of the potential, but \il{Z\simeq\mathcal{O}(1)}
elsewhere.  Indeed, this implies that in its early stages of 
evolution the axion rolls through a region with relatively large curvature,
generating a large chemical potential, and the appropriate baryon 
asymmetry is easily produced.  However, as the field falls toward the 
minimum of the potential, the enhancement \il{Z\gg 1} effectively 
``flattens'' it, suppressing the axion mass.
This enhancement also has the effect of suppressing the rate 
of axion decays to SM particles.  These two considerations taken 
together imply a sufficiently stable DM candidate that
can simultaneously generate the observed BAU.

Notably, the intermediate region of such potentials can 
give rise to highly non-trivial dynamics.
In particular, we find a period of \emph{tracking} behavior, similar to
that found in quintessence models of dark energy.  In this phase
the axion follows an attractor-like trajectory, with its equation-of-state 
parameter converging rapidly to a value that depends on the details of $Z(\phi)$
and the background cosmology.
Consequently, the axion relic abundance is rendered insensitive to 
the initial misalignment of the field, in contrast to traditional
expectations.  Furthermore, the axionic isocurvature perturbations 
also evolve in a non-trivial manner, experiencing a suppression 
in amplitude for as long as tracking continues, 
which can be a considerable duration.
The generic suppression of this isocurvature mode is one 
of several features which leads to a different 
analysis of the cosmic microwave background (CMB) 
constraints for such models.

The paper is organized as follows.  
In Sec.~\ref{sec:GeneralDescription}, we introduce a 
general model which forms the basis 
for our analysis in the remainder of the paper.  
We first discuss some of its important properties,
illustrating the non-trivial axion dynamics that arise
and producing estimates for the lifetime and relic abundance.
We then incorporate the spontaneous baryogenesis mechanism into the 
model, outlining the different avenues by which baryon asymmetry may be produced
and underscoring the significance of its interplay with the 
axion dynamics.  We also discuss the form of isocurvature
perturbations that appear in the model.
The penultimate Sec.~\ref{sec:AnExplicitModel} is devoted to an explicit 
realization, in which we demonstrate how the above scheme 
could be furnished from a complete model construction.  To this end, 
we consider an extra-dimensional ``continuum-clockwork'' model,
where our axion corresponds to the lightest state in a Kaluza-Klein (KK) tower
of axion modes.  We determine the phenomenological viability of this model and 
thereby a ``proof of concept'' showing how the ideas in this paper may be applied
within a specific setting.  Finally, in Sec.~\ref{sec:Conclusions} we 
provide a summary of our main results and possible directions for future work.

This paper also contains two Appendices.  
In Appendix~\ref{sec:TrackingDynamics} we provide a brief review of the classification 
of tracking potentials relevant for our analysis. 
Meanwhile, a derivation of the Boltzmann evolution for $(B-L)$ is
detailed in Appendix~\ref{sec:BoltzmannEquations}, which carefully accounts 
for the various subtleties of sphaleron equilibrium.

\FloatBarrier
\section{General Model Description\label{sec:GeneralDescription}}


In this section, we delineate a general model for an axion-like field which shall 
serve as the basis for our analysis in this paper.  
We begin by defining the model and examining its dynamical evolution,
and then we shift focus to incorporating a mechanism for spontaneous 
baryogenesis.  Finally, we close the section with an analysis of the 
isocurvature perturbation spectrum.

\subsection{Axion dynamics and relic abundance\label{subsec:AxionDynamics}}

Let us consider a model for an angular axion-like field $\theta(x)$ with 
periodic potential $U(\theta)$ and non-trivial wavefunction renormalization $Z(\theta)$,
such that the Lagrangian contains a non-canonical kinetic term:
\beq\label{eq:generaleffectiveL}
    \mathcal{L}_{\text{eff}}
 = \frac{f^2 }{2} Z(\theta)(\partial_\mu\theta)^2
    - \Lambda^4 U(\theta) 
    + \cdots \ .
\eeq
We have refrained from writing any topological interactions since these
will not affect our discussion of the dynamics that follow.
The two mass parameters that characterize the model are determined by UV physics. 
Namely, $f$ is the spontaneous symmetry-breaking scale associated with our
axion, and $\Lambda$ is the scale of some non-perturbative physics, \eg, the confinement 
scale of a non-Abelian gauge theory.\footnote{For example, in the case of the QCD axion, 
$f$ is associated with the scale at which 
the Peccei-Quinn $\UI_{\text{PQ}}$ symmetry is spontaneously broken, 
and $\Lambda$ with the confinement scale of QCD.}
In this paper, we shall not further address the origin of these parameters.

While a sinusoidal form \il{U(\theta)=1-\cos\theta} may be associated 
with a potential generated through instanton effects, 
the wavefunction renormalization $Z(\theta)$ is less restricted.  
A field-dependent wavefunction renormalization may arise from a variety of mechanisms,
\eg, integrating-out heavy degrees of freedom~\cite{Rubin:2001in,Tolley:2009fg,Dong:2010in}
or non-minimal couplings to gravity~\cite{Bezrukov:2007ep,Kallosh:2013tua}.
The most activity in this area has been with
inflationary model building~\cite{ArmendarizPicon:1999rj,Alishahiha:2004eh,Domcke:2017fix}
or kinetically driven quintessence~\cite{Chiba:1999ka,ArmendarizPicon:2000dh,ArmendarizPicon:2000ah}.
However, apart from some exceptions~\cite{Alvarez:2017kar}, 
the implications of these effects have not been extensively
explored in the context of axion DM models.

Any non-trivial field dependence in $Z(\theta)$ can significantly 
influence how the axion evolution unfolds, as it follows the equation of motion
\beq\label{eq:EofM_theta}
\ddot\theta
+ \frac{1}{2 Z}\frac{dZ}{d\theta}\dot\theta^2
+ 3 H \dot\theta + \frac{\Lambda^4}{f^2}\frac{1}{Z}\frac{dU}{d\theta} = 0 \ ,
\eeq
where dots indicate time derivatives $\p/\p t$. 

We can continue along these lines, analyzing the dynamics according to 
Eq.~\eqref{eq:EofM_theta}.  However, it is also instructive to introduce the 
canonically normalized field
\beq\label{eq:can_phi}
\phi(x) \equiv f\int_0^{\theta(x)} \hskip -0.2cm \sqrt{Z(\theta)}d\theta
\eeq
and study its corresponding dynamics.  In particular, the equation 
of motion takes the more familiar form
\beq\label{eq:canonicaleqnofmotion}
\ddot{\phi} + 3H\dot{\phi} + \frac{d\Veff}{d\phi} = 0 \ ,
\eeq
where \il{\Veff(\phi)\equiv \Lambda^4 U[\theta(\phi)]}, and $\theta(\phi)$
is obtained by inverting Eq.~\eqref{eq:can_phi}.
In this picture, the influence of $Z(\theta)$ is captured
solely through the \emph{deformations it induces on the canonical potential} $\Veff(\phi)$.
Naturally, in regions of field space where \il{Z\simeq\mathcal{O}(1)},
the deformation is insignificant, and $\Veff(\phi)$ is similar to the potential 
in the non-canonical representation.  However, in regions with an 
enhancement \il{Z\gg 1}, the effect is to ``flatten'' the 
canonical potential, as seen explicitly through 
\beq
\frac{d\Veff}{d\phi} = \frac{1}{\sqrt{Z}}\frac{\Lambda^4}{f}\frac{dU}{d\theta}
\eeq
and the curvature 
\beq
\frac{d^2\Veff}{d\phi^2} = 
\frac{\Lambda^4}{f^2}\frac{1}{Z}\left(\frac{d^2U}{d\theta^2} - \frac{1}{2Z}\frac{dZ}{d\theta}\frac{dU}{d\theta}\right) \ .
\eeq

In this paper, we examine the possibility that such an axion
can simultaneously (i) generate the observed BAU through 
spontaneous baryogenesis and (ii) serve as a 
DM candidate with the appropriate relic abundance.  
At first glance, the ingredients necessary to realize this appear incompatible.
Indeed, in spontaneous baryogenesis
the production of baryon asymmetry is driven by the \emph{velocity} of the 
axion field $\p_0\theta$, and thus ultimately depends on the shape of the 
potential traversed by the axion during its early evolution.  
In other words, a sufficiently steep region within $\Veff(\phi)$ is required 
for baryogenesis by these means.  With the usual sinusoidal axion 
potential, this is to equivalent requiring a sufficiently large axion mass.   
However, previous studies suggest 
this mass must be so large that the axion is rendered highly unstable
and thus an unsuitable DM candidate~\cite{Kusenko:2014uta,DeSimone:2016ofp}.  

By contrast, we argue that deformations to $\Veff(\phi)$ can repair this incompatibility.  
For the moment, we interpret the wavefunction renormalization $Z(\theta)$ 
simply as a vehicle for supplying the necessary deformations.  Then, an enhancement
\il{Z\gg 1} around the minimum of the potential --- but \il{Z = \mathcal{O}(1)} 
elsewhere --- can furnish a model with both an adequate baryon asymmetry,
as well as a suppressed axion mass and decay rates. 

To explore the implications of such a model more explicitly, 
let us consider the wavefunction
\beq\label{eq:Z(theta)}
Z(\theta) \simeq 
\left\{
    \begin{array}{lcl} 
        1 & \text{for} & \theta=\mathcal{O}(1) \\   
        1/\theta^{2n} & \text{for} & \epsilon \lesssim |\theta| < \mathcal{O}(1) \\ 
        1/\epsilon^{2n} &\text{for} & |\theta|\lesssim \epsilon 
    \end{array} \right.  \ ,
\eeq 
where \il{n>0} is an integer. Along with $n$, the small parameter \il{\epsilon > 0} 
determines the strength of the deformation.  Namely, the effective axion mass is suppressed as
\beq
\mphi^2 \equiv \left.\frac{d^2 \Veff}{d\phi^2}\right|_{\phi=0} \!\!\!\! = \epsilon^{2n}\frac{\Lambda^4}{f^2} \ .
\eeq

\begin{figure}
    \begin{center}
    \includegraphics[keepaspectratio, width=0.49\textwidth]{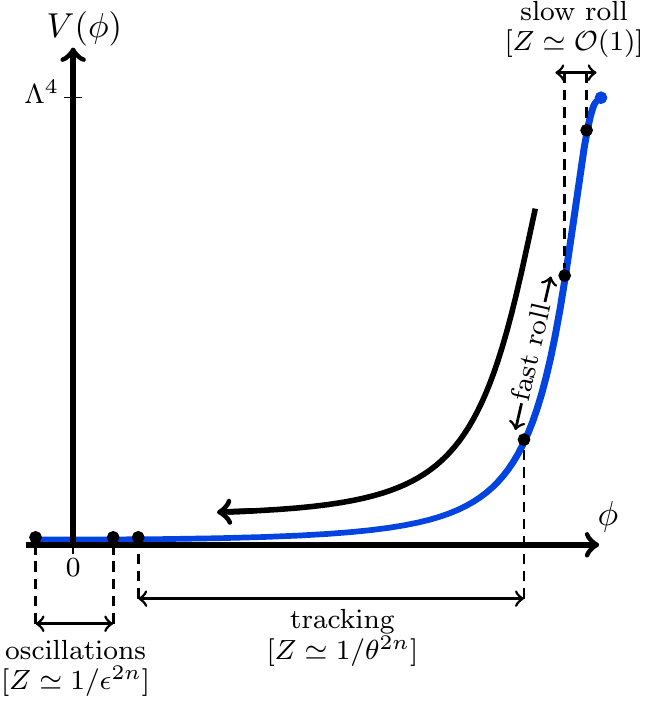}
    \end{center}
\caption{
    A schematic of the effective potential $\Veff(\phi)$ for the 
    canonically normalized field $\phi$, associated with the 
    wavefunction renormalization $Z(\theta)$ 
    in Eq.~\eqref{eq:Z(theta)}.  The non-trivial field dependence in $Z(\theta)$
    induces deformations in the potential and alters the axion dynamics.
    The periods of evolution --- slow roll, fast roll, tracking, 
    and coherent oscillations --- are labeled in 
    their respective regions.  The initial misalignment of the axion $\phiin$ is assumed
    to be toward the edge of field space.
}
\label{fig:potential_annotation}
\end{figure}

The dynamics that arises in response is generally non-trivial
and reveals trajectories qualitatively different from the traditional axion dynamics.
In the following, we outline the various periods of field evolution.
A schematic of the canonical potential $\Veff(\phi)$ is shown
in Fig.~\ref{fig:potential_annotation}, with regions labeled by their 
associated dynamics.  We shall discuss the timeline of axion field evolution, 
moving sequentially from right to left in the figure.

\subsubsection{Slow-roll and fast-roll periods\label{subsubsec:SlowRollPeriod}}

Let us assume the global symmetry associated with the axion is broken either before or
during inflation, and that the field is initially misaligned at some angle
\il{|\thetain|=\mathcal{O}(1)}.  Then, according to Eq.~\eqref{eq:Z(theta)},
our initial conditions have \il{Z \simeq 1}.
Furthermore, we shall only consider scenarios in which the axion is a light field during inflation
and the subsequent reheating epoch, such that
\beq
\frac{\Lambda^2}{f} \ll H \ ,
\eeq 
where \il{H=H(t)} is the Hubble parameter during that era.  
The damping imposed by the Hubble term holds the field to a 
slow-roll trajectory~\cite{Kawasaki:2011pd}
\beq\label{eq:slowrolltrajectory}
\dot\phi \simeq -\frac{1}{5H}\frac{d\Veff}{d\phi}
\eeq
as the radiation-dominated epoch is approached.  The slow-roll evolution
continues for as long as the following condition is satisfied:
\beq\label{eq:slowrollcondition}
\frac{1}{5H^2}\abs{\frac{d^2\Veff}{d\phi^2}} \ll 1 \ .
\eeq

The Hubble damping \il{H\sim 1/t} eventually falls sufficiently 
to violate Eq.~\eqref{eq:slowrollcondition}.  The field then enters a transient 
``fast-roll'' period in which \il{H\approx \Lambda^2/(\sqrt{5}f)}, and the 
velocity of the field reaches its maximum value over the evolution
\il{|\p_0\phi| \sim \Lambda^2/\sqrt{5Z}}.  

The temperature of the thermal bath at this point is approximately
\beq\label{eq:TFR}
T_{\text{FR}} \approx \left(\frac{18}{\pi^2 g_*}\right)^{\!\frac{1}{4}}\sqrt{\frac{\mphi M_P}{\epsilon^n}}   \ ,
\eeq
in which $g_*$ is the effective number of relativistic degrees of freedom and 
\il{M_P\equiv 1/\sqrt{8\pi G}\approx 2.4\times 10^{18}\,\GeV} is the reduced Planck scale.
The inverse dependence on the small parameter \il{\sim 1/\sqrt{\epsilon^n}} 
is particularly noteworthy, as it implies the fast-roll period is driven 
to increasingly higher temperatures as the potential is more acutely deformed.

\subsubsection{Tracking period\label{subsubsec:TrackingPeriod}}

In the conventional axion dynamics [\ie, a model with \il{Z(\theta)=1} for all $\theta$]
the field would now transit into a harmonic
region of the potential and undergo coherent oscillations. 
However, in our model, as the field enters
\il{|\theta| \lesssim \mathcal{O}(1)}, the wavefunction changes form 
to $Z(\theta)\simeq 1/\theta^{2n}$.  This change, and its associated deformation in $\Veff(\phi)$,
dramatically alters the field trajectory and introduces a starkly 
different segment of evolution.  In explicit terms, the canonical potential in this region is
\beq
\Veff(\phi) \simeq
    \frac{1}{2}\epsilon^2\Lambda^4
    \left\{
    \begin{array}{lcl}
        e^{2|\phi|/f -2}  & \text{for} & n=1  \\
        \left[{n}-(n-1)\frac{\epsilon^{n-1}}{f}|\phi|\right]^{-\frac{2}{n-1}} & \text{for} & n>1
    \end{array}\right. \ .
\label{eq:tr_pot}
\eeq

The sort of dynamics induced by such a potential is well-known in the literature of 
quintessence models~\cite{Wetterich:1987fm,Ratra:1987rm}.
In particular, $\Veff(\phi)$ yields so-called ``tracker'' solutions: 
attractor-like field trajectories 
which have an identical late-time evolution for a wide range
of initial conditions~\cite{Zlatev:1998tr,Steinhardt:1999nw}.  
These are characterized by an equation-of-state 
parameter (for axion pressure $P_{\phi}$ and energy density $\rhophi$)
\beq
\wphi \equiv \frac{P_{\phi}}{\rhophi} = \frac{\frac{1}{2}\dot{\phi}^2 - \Veff(\phi)}{\frac{1}{2}\dot{\phi}^2 + \Veff(\phi)}
\eeq
which converges to some fixed value, depending on parameters in
the potential and the background cosmology.
In Sec.~\ref{subsec:Classification} we have provided a brief overview 
of the identification and classification of tracker solutions.  
Using that technology, we deduce that tracker solutions exist 
with $\Veff(\phi)$ for \il{n \geq 1},
which drive the axion equation-of-state parameter to
\beq\label{eq:wphitrack}
w_{\phi} = \frac{1 + w - n}{n} 
\eeq
for background parameter $w$.  

In other words, $n$ determines whether the axion energy density \il{\rho_{\phi}\propto a^{-3(1+\wphi)}}
dissipates less rapidly than the dominant component in the universe, which we
assume is the radiation component (\il{w=\frac13}).
The case of an exponential potential (\il{n=1}) is unique,
since it implies $\wphi$ simply traces the background $w$ and the axion
abundance \il{\Omegaphi\equiv\rhophi/(3M_P^2H^2)} remains fixed.  For potentials with larger $n$,
the axion component behaves increasingly like vacuum energy, causing $\Omegaphi$
to grow and eventually dominate if tracking lasts sufficiently long.
Note that since we assumed $n$ is a positive integer, there is a bound \il{\wphi \leq w} 
and the axion energy density never dissipates more rapidly than the dominant component.

\subsubsection{Coherently oscillating period\label{sec:OscillationPeriod}}

The tracking dynamics continues for as long as the angular field is confined 
to the \il{\epsilon \lesssim |\theta| \lesssim \mathcal{O}(1)} region, \ie, 
for as long as the potential has the form in Eq.~\eqref{eq:tr_pot}.  
However, as the field falls to \il{|\theta|\lesssim \epsilon} it exits the 
tracking regime, and the wavefunction is frozen at a constant value \il{Z(\theta) \simeq 1/\epsilon^{2n}}.
The potential is then approximately quadratic:
\beq
\Veff(\phi)\simeq\frac{1}{2}\frac{\epsilon^{2n}\Lambda^4}{f^2}\phi^2\quad\text{for}
\quad|\phi|\lesssim\frac{f}{\epsilon^{n-1}}
\eeq
and does not support a tracker solution. 

To determine the field evolution during this final phase, we should compare
the Hubble damping $H$ to the axion mass \il{\mphi=\epsilon^n\Lambda^2/f}
at the time tracking completes.  If we find that \il{3H \lesssim \mphi} the field
will promptly begin to undergo coherent oscillations. On the other hand, if we
find \il{3H\gtrsim \mphi}, then it will sit in an overdamped phase until $H$ has dropped
sufficiently for oscillations to commence.  Assuming tracking lasts sufficiently long for 
the field to converge to the tracker trajectory, we can use Eq.~\eqref{eq:xdef}, \eqref{eq:trackingcond}, 
and \eqref{eq:wphitrack} to show that
\beq\label{eq:oscillationcond}
\frac{3H}{\mphi} \approx \sqrt{\frac{9n}{6n-4}}
\eeq
at the time the field exits the tracking region.  The ratio above is 
contained within \il{1.2\lesssim 3H/\mphi\lesssim 2.1}, so the field commences 
coherent oscillations relatively soon after, regardless of $n$.  
Neglecting the minor $n$-dependence and any ``overshooting'' effect, 
the axion energy density at the time of oscillations is given approximately by
\beq
\rho_\phi \approx \Veff(\phi) \approx  \frac{1}{2}\epsilon^{2}\Lambda^4 \ .
\label{eq:osc_den}
\eeq

Once coherent oscillations begin, the axion equation-of-state parameter 
averages to \il{\langle\wphi\rangle\rightarrow 0} and thus the axion behaves as matter.  
Ideally, the matter it constitutes would be abundant enough today
to comprise the entirety of the DM.  To calculate the relic abundance,
we use that the temperature at which oscillations first occur is 
given approximately by \il{3H\approx \mphi}:
\beq
T_{\text{osc}}^4 \approx \frac{90M_P^2\mphi^2}{g_{*}\pi^2} \ .
\eeq
Then, employing conservation of entropy density,
we find a relic abundance\footnote{
We shall often assume a temperature regime sufficiently high for this dynamics 
that the effective relativistic $g_*$ and entropy degrees of freedom $g_{*S}$ are 
approximately equal.} 
\beq\label{eq:general_abundance}
\Omegaphi h^2 \approx \frac{\epsilon^{2(1-n)}}{g_{*}^{1/4}}\sqrt{\frac{\mphi}{7\,\eV}}\left(\frac{f}{10^{12}\,\GeV}\right)^2
\eeq 
for a sufficiently long-lived axion.  

It is important to note that for \il{n\neq 1} a deformation in the potential
enhances the relic abundance, while for \il{n=1} it is \emph{independent of $\epsilon$}.
Moreover, $\Omegaphi$ is \emph{independent of the initial misalignment angle $\thetain$},
as a result of the tracking dynamics encountered in the field evolution.
This insensitivity reveals a significant departure from the traditional axion cosmology
and also leads to a natural suppression of axionic isocurvature perturbations. 
We shall discuss the perturbations in more detail below and also provide 
a model-specific analysis in Sec.~\ref{sec:AnExplicitModel}.  

Meanwhile, the lifetime of the axion is also enhanced
if it decays primarily through an anomalous coupling to the electroweak sector.
That is, the enhancement in $Z(\theta)$ near the origin generically implies
a decay width \il{\Gamma_{\phi}\propto \mphi^3/[f^2Z(0)]} and thus an enhanced lifetime 
\beq
    \tau_{\phi} \approx \frac{6.6\times 10^{30}\,\s}{\epsilon^{2n}} \left(\frac{f}{10^{12}\,\GeV}\right)^2 \left(\frac{7\,\eV}{\mphi}\right)^3  \ ,
\eeq
alleviating constraints from axion decays as the potential is deformed with \il{\epsilon\ll 1}.

\subsection{Incorporating the baryogenesis mechanism\label{subsec:IncorporatingBaryogenesis}}

Let us now shift focus to embedding the mechanism for spontaneous baryogenesis in our model.
We shall begin by describing the interactions necessary and discussing how they may appear
in the UV theory.  We then construct the Boltzmann equations for the
matter-antimatter asymmetry and catalog the different ways in which production can occur.

\subsubsection{Spontaneous \texorpdfstring{$CPT$}{CPT} violation}

We must include interactions that spontaneously violate $CPT$ in the
axion background, such as the effective coupling between 
the axion $\theta$ and the baryon current $J_B^{\mu}$:
\beq\label{eq:requiredterm}
\mathcal{L}_{\text{eff}} \supset \frac{1}{\mathcal{N}} \, \p_{\mu}\theta \cdot J_{B }^{\mu}
\eeq
where $\mathcal{N}$ is a constant that we clarify in what follows.
The baryon current is given by
\beq
    J^{\mu}_{B }=\frac{1}{3}\sum_k \left(q^{\dagger}_k\overline{\sigma}^{\mu}q_k
    +u^*_k\overline{\sigma}^{\mu}u_k +  d^{*}_k\overline{\sigma}^{\mu}d_k
    \right)\ ,
\eeq
where $q_k$, $u_k$, and $d_k$ are two-component Weyl spinors for the
left-handed quark doublets, right-handed up-type quarks, and 
right-handed down-type quarks, respectively.
Note also that we have suppressed $\SUIII_c$ and $\SUII_L$ indices
and the $\sigma^{\mu}$ are the Pauli matrices. 

The homogeneity of the axion field in space implies its gradient is negligible and 
that Eq.~\eqref{eq:requiredterm} effectively reduces to an interaction
\beq\label{eq:effectiverequiredterm}
\mathcal{L}_{\text{eff}} \supset \frac{1}{\mathcal{N}}\,\p_0 \theta \cdot J_B^{0} \ .
\eeq
Indeed, such a term spontaneously breaks $CPT$ symmetry once the field 
is set in motion, inducing an energy gap between baryons and antibaryons. 
If \mbox{$B$-violating} interactions are occurring in the thermal bath, 
then a baryon asymmetry is generated.
 
There are several ways to motivate the appearance of an interaction 
such as Eq.~\eqref{eq:requiredterm} in the effective Lagrangian.  
For instance, we may consider a spontaneous breaking of the
baryon number symmetry $\UI_B$ at high scales, in which $\theta$ is the corresponding
Nambu-Goldstone boson (NGB), which in general would appear as 
the phase of some complex scalar field.  In such a scenario, it is also 
necessary to specify the relationship 
between the axion potential and $\UI_B$, as well as the effect of \mbox{$B$-violating}
operators after integrating-out the radial scalar field. 

However, other avenues exist through which we may generate such an 
interaction, even if $\theta$ is neutral under $\UI_B$.
By rotating the quark phases according to
\begin{align}
q_k &\longrightarrow e^{i \theta/\mathcal{N}}q_k \nn \\
u_k &\longrightarrow e^{i \theta/\mathcal{N}}u_k \nn \\
d_k &\longrightarrow e^{i \theta/\mathcal{N}}d_k \ ,
\end{align}
our term in Eq.~\eqref{eq:requiredterm} can be eliminated from the action,
and we obtain an equivalent operator
\beq
\label{eq:CSterm}
\mathcal{L}_{\text{eff}} \supset \frac{3}{\mathcal{N}}\,\theta\left[\frac{\alpha_2}{16\pi}W^{\mu\nu}\Wt_{\mu\nu}-
\frac{\alpha_Y}{8\pi}B^{\mu\nu}\Bt_{\mu\nu}\right] \ ,
\eeq
with $W_{\mu\nu}$ the $\text{SU}(2)_L$ field strength, $B_{\mu\nu}$ the $\text{U}(1)_Y$ field strength, and
$\Wt_{\mu\nu}$ and $\Bt_{\mu\nu}$ their respective duals.
The factors \il{\alpha_2=g_2^2/(4\pi)} and \il{\alpha_Y=g_Y^2/(4\pi)} are the 
weak and hypercharge gauge coupling constants, respectively.
Such an axion-like coupling of $\theta$ to the gauge bosons can naturally arise 
in the UV theory, independent of baryon number~\cite{
    Witten:1984dg,
    Green:1984sg,
    Choi:1985je,
    Witten:1985fp,
    Choi:1985bz,
    Svrcek:2006yi},
if the normalization of the interaction is given by
\beq
\mathcal{N} = 3/n \quad \text{for} \quad n=1,2,\cdots \ .
\eeq
In what follows, we shall take \il{n=1} for simplicity.

\subsubsection{Source for \texorpdfstring{$\ilBmL$}{B-L} violation}

In addition to spontaneous $CPT$-violation, to satisfy the conditions
for baryogenesis some \mbox{$B$-violating} interactions must also exist.
The early-universe plasma is naturally equipped with such processes 
through weak sphaleron transitions.  However, since the weak sphalerons
preserve $(B-L)$ any baryon number generated through the spontaneous baryogenesis 
mechanism will still be annihilated once the axion settles to its minimum vacuum value.
It is therefore essential that our theory also include interactions which 
break $(B-L)$.  A well-motivated way to invoke such terms is through physics in the neutrino sector,
where heavy right-handed neutrinos offer a natural explanation of neutrino masses 
and other associated phenomena.  In the low-energy theory these appear in the
form of a Weinberg operator
\beq
\mathcal{L}_{{L\hskip -0.15cm\slash}} = \frac{(\ell_i \Higgs)(\ell_j \Higgs)}{2M_*} \ ,
\label{eq:wein_op}
\eeq
where $\ell_i$ is an SM lepton doublet, $\Higgs$ is the Higgs doublet,
and neutrino observables determine the mass scale \il{M_*\simeq (10^{14} - 10^{15})\,\GeV}.
Of course, the Weinberg operator breaks the lepton number $L$
in addition to $(B-L)$, so we shall denote associated processes by $\LV$.
The existence of heavy right-handed neutrinos in the early universe
may also lead to successful thermal leptogenesis through
out-of-equilibrium decays~\cite{Fukugita:1986hr}.  However, 
our study is dissociated from these models since $M_*$ is sufficiently heavy that
the right-handed neutrinos are never produced in the thermal bath.
The operator in Eq.~\eqref{eq:wein_op} thus remains valid throughout our analysis.

\subsubsection{Boltzmann evolution of asymmetry}

Under the spontaneous breaking of $CPT$ by the term in Eq.~\eqref{eq:effectiverequiredterm},
one may readily show that for sufficiently rapid \mbox{$(B-L)$-violating} processes, we obtain an
equilibrium number-density asymmetry
\beq
\label{eq:nBmLeq}
    n^{\text{eq}}_\BmL = \frac{1}{6}\muBmL T^2\left[1+\mathcal{O}\left(\frac{\muBmL}{T}\right)^2\right] \ ,
\eeq
where $\muBmL$ is the effective chemical potential associated with $\ilBmL$.
However, the asymmetry is not necessarily generated at equilibrium;
rather, its evolution is more generally described by the Boltzmann equation
\beq\label{eq:lep_evol}
\dot n_\BmL + 3 H n_\BmL =  -\GammaLV  (n_\BmL - n_\BmL^{\text{eq}}) \ .
\eeq
A detailed derivation of the Boltzmann equation, in which we account for the role
of sphaleron transitions in the plasma, is provided in Appendix~\ref{sec:BoltzmannEquations}.
Implementing these we can extract the interaction rate 
\beq
    \GammaLV= \frac{9(171+65 N_f - 6N_f^2)}{45+73 N_f -3 N_f^2}\cdot\frac{\gammaLV}{T^3} \ ,
\eeq
in which
\beq\label{eq:gammaLVdef}
\gammaLV = \mathcal{O}(0.01) \frac{T^6}{M_*^2}
\eeq
is the thermally averaged scattering rate density for processes sourced 
by the Weinberg operator in Eq.~\eqref{eq:wein_op}, and $N_f$ denotes the number of 
generations with Yukawa interactions in equilibrium during baryogenesis. 
The effective chemical potential is likewise given by
\beq\label{eq:muBmL}
\muBmL = -g(N_f)\cdot \partial_0 \theta \ ,
\eeq
for which the coefficient is derived in Eq.~\eqref{eq:nBmLeqapp}:
\beq
g(N_f) = \frac{4}{3}\cdot\frac{36+65 N_f -6 N_f^2}{171+65 N_f -6 N_f^2} \ .
\eeq
The weak sphaleron processes eventually decouple, and the final baryon number is set according to
\beq\label{eq:sphaleronfactor}
n_B =\frac{28}{79} n_\BmL \ .
\eeq

The characteristic scale which determines the final $n_\BmL$, however, is the temperature
$\TD$ at which the processes derived from $\mathcal{L}_\LV$ decouple.  In explicit terms,
we define $\TD$ as the temperature at which the Hubble rate becomes dominant 
\il{H \geq \GammaLV}.  It is expressed as
\beq
    \TD \simeq \mathcal{O}(\text{few})\sqrt{g_*}\,\frac{M_*^2}{M_P} \ ,
\eeq
which in our study will be given by \il{\TD\approx 10^{13}\,\GeV}.
The conclusion we draw is that, given the assumptions above, a relatively high reheating 
temperature \il{\TRH \gtrsim \TD} is necessary for processes
sourced by Eq.~\eqref{eq:wein_op} to ever achieve equilibrium. 
Indeed, depending on the temperature regime of the early radiation-dominated epoch, 
there are several different ways in which baryogenesis may unfold.
To explore these in more detail, let us work instead with comoving
quantities, such as the abundance \il{Y_\BmL\equiv n_\BmL/s}, in which
\beq\label{eq:entropydensity}
s = \frac{2\pi^2}{45}g_{*S}T^3 
\eeq
is the entropy density.  It is then straightforward to 
determine the corresponding Boltzmann evolution 
\beq
\label{eq:YBmLboltzmann}
\frac{dY_\BmL}{dT} = \frac{Y_\BmL - Y_\BmL^{\text{eq}}}{\TD} \ ,
\eeq
where the equilibrium value \il{Y_\BmL^{\text{eq}}\equiv n_\BmL^{\text{eq}}/s} is defined analogously to Eq.~\eqref{eq:nBmLeq}.
An integral solution follows as
\begin{align}
\label{eq:lep_yield} 
    Y_\BmL(T) = \int_T^{\TRH} dT' \left[\frac{e^{-(T'-T)/\TD} }{\TD}\right] Y_\BmL^{\text{eq}}(T')  \ .
\end{align}
There are different limiting behaviors, depending on the relative size of the temperatures
$\TD$ and $\TRH$, corresponding to equilibrium and out-of-equilibrium production.  
Below, we discuss each of these cases.

\subsubsection{Equilibrium production\label{subsubsec:EquilibriumProduction}}

In the \il{\TD\ll \TRH} regime, the function enclosed within square brackets of
Eq.~\eqref{eq:lep_yield} approaches a Dirac delta function \il{\delta(T'-T)},
which holds for temperatures \il{T > \TD}.
The asymmetry closely follows its equilibrium value \il{Y_\BmL(T) \simeq Y_\BmL^{\text{eq}}(T)}
in that regime and reproduces the result in Eq.~\eqref{eq:nBmLeq}.
However, once the plasma cools below \il{T < \TD}, equilibrium productions ceases
and the asymmetry ``freezes out.'' Therefore, we obtain a late-time asymmetry 
\beq\label{eq:FOlepasy}
    Y_\BmL \simeq Y_\BmL^{\text{eq}}(\TD) \ .
\eeq

The interplay between the equilibrium production of $Y_\BmL$ and the axion dynamics is
significant in determining the final baryon asymmetry.
In particular, the late-time yield can be ruined if decoupling does not 
occur until after the axion undergoes oscillations --- in this event $\mu_\BmL$ 
oscillates as well, and the asymmetry is washed out.
Almost as severely, if decoupling occurs within the tracking period,
the field velocity and thus $Y_\BmL$ is considerably weakened.  In the 
context of equilibrium production we shall then assume that
\beq\label{eq:washoutcondition}
\frac{\Lambda^2}{f} \ll \sqrt{5}H(\TD) \approx 3 \TD^2/M_P \ ,
\eeq
which implies the field is slowly rolling at decoupling.  
The yield is then determined by the slow-roll trajectory 
in Eq.~\eqref{eq:slowrolltrajectory} and we find an approximate expression
\beq
    Y_\BmL \sim 10^{-3}\frac{1}{\epsilon^{2n}}\frac{\mphi^2M_P}{\TD^3} \ .
\eeq
Note that taking $\epsilon$ to smaller values, and thereby deforming the 
potential more acutely, corresponds to an \emph{enhancement} in 
the matter-antimatter asymmetry.

\subsubsection{Out-of-equilibrium production\label{subsubsec:OutofEquilibriumProduction}}

Let us now consider the opposite temperature regime, 
in which reheating occurs below the scale of decoupling \il{\TRH < \TD}. 
In such a scenario, processes which violate $\ilBmL$ are always out-of-equilibrium,
\ie, \il{Y_\BmL < Y_\BmL^{\text{eq}}} for all temperatures.
Consequently, the asymmetry production occurs in a different
fashion, weakly but persistently driven by the \il{Y_\BmL^{\text{eq}}/\TD} term in 
Eq.~\eqref{eq:YBmLboltzmann}.  The production mechanism here is 
analogous to the ``freeze-in'' production in the DM literature~\cite{Hall:2009bx}.  
It is straightforward to obtain the yield
\beq\label{eq:yield_fi}
Y_\BmL \simeq \frac{32}{25\pi^2g_*} \frac{\Lambda^2}{f \TD} \mathcal{I}(\thetain) \ ,
\eeq
in which the integral $\mathcal{I}(\theta)$ is defined by
\beq\label{eq:Iintegral}
\mathcal{I}(\theta) \equiv \int_0^{\infty} du \,\frac{1}{u} \frac{d\theta}{du} \ ,
\eeq
for a dimensionless temporal parameter \il{u\equiv\Lambda^2t/f}.
It can be shown using Eq.~\eqref{eq:EofM_theta} that \il{\mathcal{I}(\thetain)} gives
only an $\mathcal{O}(1)$ contribution.
Finally, assuming baryogenesis occurs at temperatures \il{T \gtrsim 100\, \GeV}, 
we conclude that
\beq
    Y_\BmL \sim 10^{-3}\frac{\Lambda^2}{f \TD} = 10^{-3} \frac{1}{\epsilon^n}\frac{\mphi}{\TD}
\label{eq:FIlepasy}
\eeq
within order-of-magnitude accuracy.  

We observe for both the equilibrium production in Eq.~\eqref{eq:FOlepasy}
and the out-of-equilibrium production above, for a fixed axion mass
\emph{the asymmetry is enhanced by deformations \il{\epsilon\ll 1} in the potential}.  
Thus, possibilities for model-building can exist in either of these two regimes.  

On another note, by comparing the scaling behavior for the axion 
relic abundance \il{\Omegaphi\propto \epsilon^{2(1-n)}}
[see Eq.~\eqref{eq:general_abundance}] to the estimates 
for the late-time asymmetry $Y_\BmL$ above,
we find that \il{n=1} has some intriguing properties.
In particular, while the baryon asymmetry is always 
enhanced by \il{\epsilon \ll 1}, the relic abundance is 
unaffected for \il{n=1}, providing us with some modularity between these 
two cosmological quantities.  We shall investigate these details further in
the context of the explicit model constructed in Sec.~\ref{sec:AnExplicitModel}, 
which is specific to the \il{n=1} case.

\subsection{Isocurvature perturbations\label{subsec:IsocurvaturePerturbations}}

As alluded to above, for successful baryogenesis
the axion must be relatively light \il{\Lambda^2/f \ll H_I} during inflation,
for an inflationary Hubble scale $H_I$.
Therefore, it is subject to quantum fluctuations with amplitude
\beq\label{eq:axionfluctuations}
\dphi_{\text{in}} = \frac{H_I}{2\pi}  \ ,
\eeq
in the pure-de Sitter limit.  As a result, we find corresponding fluctuations in the 
angular field
\beq
    \delta\thetain = \frac{1}{\sqrt{Z(\thetain)}}\frac{H_I}{2\pi f} \ .
\eeq

The sources of primordial scalar perturbations are decomposed into 
linearly independent adiabatic and isocurvature modes, 
\ie, perturbations to the total energy density and the local 
equation of state, respectively.
The fluctuations $\delta\thetain$ source only the isocurvature mode, 
which is subdominant and tightly constrained 
by observations of the CMB~\cite{Akrami:2018odb}.  
Furthermore, since the baryon asymmetry is generated via the effective chemical potential
\il{\muBmL\sim\p_0\theta}, baryonic isocurvature perturbations $\delta Y_B$ 
also exist and play an important role.

As illustrated in Sec.~\ref{subsec:AxionDynamics}, the presence of a tracking region in 
the canonical potential $\Veff(\phi)$ renders the late-time axion dynamics insensitive 
to the initial field displacement.  Consequently, the axionic isocurvature 
perturbations are generically suppressed with a magnitude corresponding to the duration of
the tracking period.\footnote{
In the explicit model realization covered in Sec.~\ref{subsec:IsocurvaturePerturbations},
we give a more rigorous illustration of this phenomenon, supplemented by a full numerical simulation
of the system of perturbations.
}
In the remainder of this section, we shall assume tracking lasts 
for a sufficiently long period that we may focus exclusively on the baryonic component.  

As discussed above, $Y_B$ may be populated while driving the system 
either at equilibrium (freeze-out) or out-of-equilibrium (freeze-in).
In the former case, the baryon asymmetry produced is simply \il{Y_B\approx Y_B^{\rm eq}\propto \muBmL}
evaluated at the decoupling temperature $\TD$.  Therefore, the baryonic perturbation is
\begin{align}
\frac{\delta Y_B}{Y_B} \simeq \frac{\delta \muBmL}{\muBmL} 
    &= \left.\frac{d \log\mu_{B-L}^{\rm eq}}{d\theta} \frac{1}{\sqrt{Z(\theta)}}\frac{H_I}{2\pi f}\right|_{\thetain}  \nn \\
    &= \left.\frac{d\log\dot\theta}{d\theta}\frac{1}{\sqrt{Z(\theta)}}\frac{H_I}{2\pi f}\right|_{\thetain} \ ,
\end{align}
also evaluated at $\TD$.
As before, we assume decoupling occurs during the slow-roll period, 
so the trajectory is given by Eq.~\eqref{eq:slowrolltrajectory}:
\begin{align}
    \dot\theta\simeq -\frac{1}{5H}\frac{\Lambda^4}{f^2}\frac{1}{Z(\theta)}\frac{dU}{d\theta} \ ,
\end{align} 
which we have written in the non-canonical basis.  Assuming the field moves negligibly from
its initial misalignment \il{\theta\simeq\thetain}, the approximate isocurvature is
\beq\label{eq:PSSFO}
\mathcal{P}_{\mathcal{S}\mathcal{S}} \simeq 
\frac{1}{Z(\theta)}\!
\left(\!\frac{\Omega_B}{\OmegaCDM}\frac{H_I}{2\pi f}\!\right)^{\!2}\!
\left.\left[\frac{U''(\theta)}{U(\theta)} - \frac{Z'(\theta)}{Z(\theta)}\right]^{\!2}\right|_{\thetain} \!
\eeq
where a prime denotes a derivative with respect to the field $\theta$.
Interestingly, the two terms in Eq.~\eqref{eq:PSSFO} may have opposite 
signs, allowing for cancellations and a vanishing perturbation.  Additionally,
the first term can clearly vanish if $\thetain$ sits at any of its inflection points.

On the other hand, in the case of out-of-equilibrium production, we 
can derive the baryonic perturbation directly from Eq.~\eqref{eq:yield_fi}:
\beq
\frac{\delta Y_B}{Y_B} \simeq
 \left.\frac{d\log \mathcal{I}(\theta)}{d\theta} \frac{1}{\sqrt{Z(\theta)}}\frac{H_I}{2\pi f}\right|_{\thetain}  ,
\eeq
which results in the expression for the power
\beq\label{eq:PSSFI}
P_{\mathcal{S}\mathcal{S}} \simeq \left.\frac{1}{Z(\theta)}\left(\!\frac{\Omega_B}{\OmegaCDM}\frac{H_I}{2\pi f}\!\right)^{\!2}\left(\frac{d\log\mathcal{I}(\theta)}{d\theta}\right)^{\!2}\right|_{\thetain} . \\
\eeq
The integral $\mathcal{I}(\thetain)$ is over time [see Eq.~\eqref{eq:Iintegral}] and 
therefore it is not a simple function of the potential or its derivatives.  Instead, these results
must be obtained numerically from the equation of motion. 

The power spectra in Eq.~\eqref{eq:PSSFO} and Eq.~\eqref{eq:PSSFI} act as constraints 
on the parameter space given an explicit model realization.
In the remainder of this paper, we shall consider such a concrete model, and show a 
more detailed study of its phenomenology and cosmological constraints, using numerical
simulations where necessary.

\FloatBarrier
\section{An explicit model: The continuum-clockwork axion\label{sec:AnExplicitModel}}


In Sec.~\ref{sec:GeneralDescription}, we described a general 
construction by which an axion-like field $\theta$ may dynamically generate
matter-antimatter asymmetry in the early universe, while also serving 
as a plausible DM candidate.  The crux of this approach is the appearance 
of a field-dependent wavefunction renormalization $Z(\theta)$ 
that meets some basic requirements.  In particular, if the wavefunction is
enhanced \il{Z\gg 1} near the minimum of the axion potential, 
but remains \il{Z=\mathcal{O}(1)} elsewhere, then the potential $\Veff(\phi)$ for
the canonically normalized field $\phi$ is ``deformed'' in a way that suppresses its mass, generically 
suppresses its couplings, and can dramatically alter its dynamics.  

In this section, we demonstrate that models exist with the ingredients necessary to 
furnish such a wavefunction renormalization.  As an explicit example, we focus
on ``continuum-clockwork'' (CCW)~\cite{Giudice:2016yja,Craig:2017cda,Giudice:2017suc,Choi:2017ncj}
axion models\footnote{In the formalism below we rely heavily on Ref.~\cite{Choi:2017ncj}.},
incorporating the interactions necessary for spontaneous baryogenesis
along the lines of Sec.~\ref{subsec:IncorporatingBaryogenesis}.
We show that regions of parameter space exist in which both the observed baryon 
asymmetry and dark matter abundance are produced.  Furthermore, we show that 
other phenomenological constraints, such as those from decays and isocurvature perturbations, 
are adequately contained.

\subsection{Overall features and construction\label{subsec:OverallFeaturesandConstruction}}

The hallmark of the ``clockwork mechanism''~\cite{Kim:2004rp,Choi:2014rja,Choi:2015fiu,Kaplan:2015fuy}
is the generation of an exponential hierarchy of couplings in theories with exclusively $\mathcal{O}(1)$ input parameters.  
There have been many studies and implementations,
including on the QCD axion~\cite{
    Higaki:2015jag,
    Higaki:2016yqk,
    Higaki:2016jjh,
    Farina:2016tgd,
    Coy:2017yex,
    Long:2018nsl},
dark matter~\cite{Hambye:2016qkf,Kim:2017mtc,Kim:2018xsp,Goudelis:2018xqi},
cosmological 
topics~\cite{
    Fonseca:2016eoo,
    Saraswat:2016eaz,
    Kehagias:2016kzt,
    Park:2018kst,
    Agrawal:2018mkd},
flavor physics~\cite{
    Hong:2017tel,
    Park:2017yrn,
    Ibarra:2017tju,
    Alonso:2018bcg},
and generalizations or more formal
aspects~\cite{
    Giudice:2016yja,
    Craig:2017cda,
    Giudice:2017suc,
    Choi:2017ncj,
    Ahmed:2016viu,
    Ben-Dayan:2017rvr,
    Lee:2017fin,
    Giudice:2017fmj,
    Choi:2017ncj,
    Teresi:2018eai}.

To introduce this idea more explicitly, let us consider a 
model with \il{N+1} scalars $\chi_i$.
The clockwork mechanism typically arises through ``nearest-neighbor'' interactions 
between adjacent scalars, such as through terms proportional to \il{(\chi_{i+1}-q\chi_i)^2}, where
\il{q>1} is a dimensionless parameter. Then, the lightest mass eigenstate $\phi$ 
in the system exhibits most of the interesting phenomenology.  
In particular, if the $\chi_i$ have couplings $Q_i$ to some other sector,
then these contribute to the coupling for $\phi$ as
\beq\label{eq:Q_phi}
Q_{\phi} ~\propto~ \sum_{i=0}^N \frac{Q_i}{q^{N-i}} \ .
\eeq
That is, the coupling for the lightest state is determined through a 
non-uniform distribution over the $Q_i$.  As their contributions are weighted 
by powers \il{1/q^{N-i}}, the distribution is effectively ``localized'' toward $Q_N$, 
where the parameter $q$ sets the strength of the localization.

A natural extension of this idea is to construct the clockwork mechanism in the
\emph{continuum} limit \il{N\rightarrow \infty}, where the theory is reinterpreted
as that of a discretized compact extra dimension.  In the continuum, the nearest-neighbor 
interactions composed of \il{\chi_{i+1}-q\chi_i} are mapped 
onto \il{\partial_y\chi(x,y) - m\chi(x,y)}, where \il{m>0},
the extra spatial coordinate is $y$, and $\chi(x,y)$ is now a five-dimensional scalar field.
This type of combination can be realized by bulk and boundary mass terms.  
Moreover, several of the phenomena found in the discrete clockwork theory are mapped onto
the extra-dimensional theory in some way. In particular, the profile for the 
zero-mode is exponentially localized toward a boundary 
in the extra dimension, analogous to the localization of the coupling 
in Eq.~\eqref{eq:Q_phi}.  Similar phenomenological implications arise from
this as well, such as the suppression of couplings to other boundary operators.
 
Furthermore, interesting observations can be made if the CCW theory is 
constructed from a five-dimensional angular field $\theta(x,y)$.
Due to the periodicity \il{\theta\rightarrow \theta+2\pi}, 
the clockwork interactions should have the general form \il{\partial_y\theta(x,y)-m V(\theta)},
for periodic \il{V(\theta)=V(\theta+ 2\pi)}.  
As discussed in Ref.~\cite{Choi:2017ncj}, this results in a more non-trivial 
localization of the lightest mode along the extra dimension, which has subtle 
implications for the axion couplings and its dynamics.  While the specific
details are beyond the scope of this paper, it provides us
the essential features by which we shall realize a field-dependent wavefunction
renormalization of the form proposed in Sec.~\ref{sec:GeneralDescription}.

Let us therefore begin by considering the
action for this particular five-dimensional realization:
\beq\label{eq:Stheta}
\mathcal{S}_{\theta} = 
\frac{f_5^3}{2} \int d^4x dy \left[(\p_{\mu} \theta)^2 - \left(\p_y\theta - m\sin\theta\right)^2\right] \ ,
\eeq
where we compactify over an \il{\mathcal{S}_1/\mathbb{Z}_2} orbifold
of radius $R$, and $m$ sets the scale for bulk and boundary terms.
The SM fields are assumed to be confined to the \il{y=0} brane and
flat space is assumed for tractability.  
Note that a massless four-dimensional mode $\phi(x)$ is found in the spectrum:
\beq\label{eq:tantheta}
\tan\left[\frac{\theta(x,y)}{2}\right] = e^{my}u\!\left[\phi(x)\right] \ ,
\eeq
where the function $u[\phi]$ enforces canonical normalization for $\phi$
over its domain.  Namely, we define
\beq
u\!\left[\phi\right] \equiv e^{-\pi mR} \Jsc\bigg[\frac{\phi}{2f}\bigg|1 - e^{-2\pi mR}\bigg]
\eeq
in which \il{f^2\equiv f_5^3(1-e^{-2\pi mR})/(2m)} and \il{\Jsc[\,\cdot\,\vert \,\cdot\,]} is a 
Jacobi elliptic function.  Integrating out the higher KK modes and the compact dimension,
we construct the low-energy effective action 
\beq
\label{eq:ccw_wave}
    S_{\theta} \approx \frac{1}{2}\int d^4x \, \frac{f_5^3}{m}\frac{\left(\p_{\mu}\theta\right)^2}{\coth\left(\pi mR\right) - \cos\theta}  \ ,
\eeq
where it is understood that \il{\theta\equiv\theta(x,0)} is evaluated at the \il{y=0} brane.  
The wavefunction renormalization as defined in Eq.~\eqref{eq:generaleffectiveL}
is easily extracted:
\beq
    Z(\theta) \simeq \frac{1}{\coth\left(\pi mR\right) - \cos\theta}  \ .
\eeq
In the regime \il{mR\gtrsim \mathcal{O}(\text{few})} that clockwork has a substantial 
effect, this reduces to
\beq\label{eq:Zclockwork}
Z(\theta) \simeq \frac{2}{1+2\epsilon^2 - \cos\theta} \ ,
\eeq
where \il{\epsilon \equiv e^{-\pi mR}} is a small parameter.  It is manifest 
that \il{Z(\theta) \simeq \mathcal{O}(1)} near the boundaries of field space
and \il{Z(\theta) \gg 1} near the origin.  
Furthermore, expanding about the origin we find \il{Z(\theta)\approx 4/\theta^2}, which 
is the necessary scaling to ensure the desired ``tracking'' dynamics.
It is then evident that the CCW axion reproduces the \il{n=1} form of Eq.~\eqref{eq:Z(theta)}
and we can conclude: \emph{CCW axions satisfy our minimal requirements for 
spontaneous baryogenesis driven by a stable axion DM candidate.}

\begin{figure}
    \begin{center}
    \includegraphics[keepaspectratio, width=0.49\textwidth]{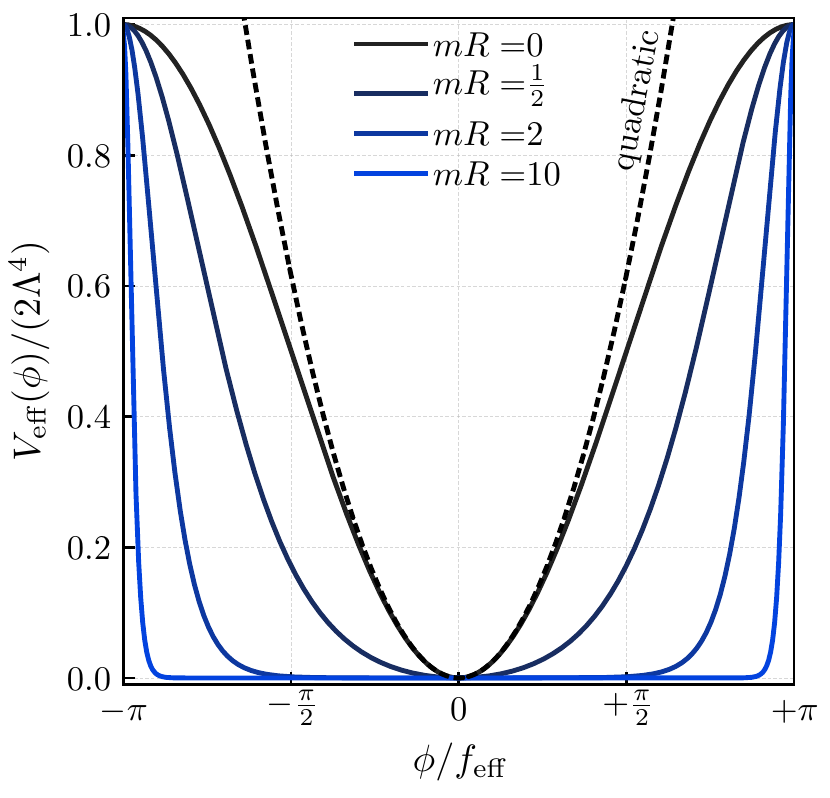}
    \end{center}
    \caption{The effective 4D canonical potential $\Veff(\phi)$ that arises for 
    a continuum-clockwork axion, where the different curves show various choices 
    for the clockwork parameter $mR$, and the horizontal axis is normalized 
    such that curves are plotted over the full field range \il{\phi\in \left[-\pi \feff, +\pi \feff\right]}.
}
\label{fig:potential}
\end{figure}

Having satisfied the minimal set of conditions from Sec.~\ref{sec:GeneralDescription},
let us further investigate the details of this model.  While the symmetry of 
the action in Eq.~\eqref{eq:Stheta} yields a massless zero-mode $\phi$, any small deviation in 
boundary masses will generate an effective four-dimensional 
potential \il{\Lambda^4(1-\cos\theta)}, which in the canonical basis reads
\beq\label{eq:Veff}
    \Veff(\phi) = \frac{2\Lambda^4}{\displaystyle{1 + \frac{1}{u^2\!\left[\phi\right]}}} \ .
\eeq
While $\Veff(\phi)$ is periodic in $\phi$, it is important 
to note the period is \emph{not} given by $2\pi f$, but rather by the expression
\begin{align}\label{eq:feff}
    2\pi f_{\text{eff}} &\equiv 4f K\left(1 - \epsilon^2\right) \nn \\
    &\xrightarrow[~\epsilon^2\ll 1~]{} 2f\log\left(\frac{16}{\epsilon^2}\right) \ ,
\end{align}
where $K(\cdot)$ is the complete elliptic integral of the first kind.
As a result, the field range of the canonical four-dimensional axion is effectively
extended for finite $mR$.

In Fig.~\ref{fig:potential} the potential is shown
for several values of $mR$, normalized so that the curves all span the same domain.
As soon as we exceed \il{mR \gtrsim \mathcal{O}(\text{few})} the potential quickly shows 
substantial deformations, with the minimum flattened
along most of the field range.  The resulting axion mass
\beq\label{eq:mphi}
\mphi^2 \equiv \!\left.\frac{\p^2\Veff}{\p\phi^2}\right|_{\phi=0} \!\!\!\! =~ e^{-2\pi mR}\frac{\Lambda^4}{f^2} \ ,
\eeq
is \emph{exponentially suppressed} relative to the standard \il{mR = 0} sinusoidal 
potential.  Indeed, the suppression of this mass scale confirms the CCW axion model 
is equipped with one of the imperative features.

The other feature necessary to avoid phenomenological complications
is the suppression of axion decays to SM states.
In the discrete clockwork theory described above Eq.~\eqref{eq:Q_phi}, 
this suppression would arise for the light state $\phi$ if the SM were coupled 
to the endpoint $\chi_N$ opposite to where $\phi$ 
is localized.  Analogously, in the continuum limit
this corresponds to SM fields confined to the \il{y=0} brane.
In other words, if we have an interaction between the 
axion and some generic SM operator $\mathcal{O}(x)$
\beq
\mathcal{S}_{\theta} \supset \int d^4x \theta(x,0)\mathcal{O}(x) \ ,
\eeq
we can write it in the canonical basis using
\beq\label{eq:fieldmapping}
\phi = f\int\!\!\sqrt{Z(\theta)}d\theta ~\approx~ 
\frac{\pi\feff}{\epsilon}\frac{F\big(\frac{\theta}{2}\big|\!-\frac{1}{\epsilon^2}\big)}{K(1-\epsilon^2)} \ ,
\eeq
where $F(\cdot\,\vert \,\cdot)$ is the incomplete elliptic integral of the first kind.  
Let us examine how the coupling is affected for larger $mR$.
Away from the minimum of the potential \il{\phi \approx \pi\feff} is quickly approached
and thus the coupling is not significantly affected beyond the minor enhancement of $\feff$. 
However, around the minimum Eq.~\eqref{eq:fieldmapping}
reduces to \il{\theta \approx \epsilon \phi/f},
so that couplings to $\phi$ are exponentially suppressed. For example, if we take
the operator \il{\mathcal{O}(x) \sim F_{\mu\nu}\Ft^{\mu\nu}}, for
some SM field strength $F_{\mu\nu}$ and its dual $\Ft_{\mu\nu}$,
then the axion decay rate suffers a suppression
\beq\label{eq:decaysuppression}
\Gamma_{\phi} \propto \frac{1}{Z(0)}\frac{\mphi^3}{f^2} = \epsilon^2 \cdot \frac{\mphi^3}{f^2}  \ .
\eeq
Note that this suppression acts in addition to that implicitly included in the mass [see Eq.~\eqref{eq:mphi}].

\begin{figure*}[t]
    \begin{center}
    \includegraphics[keepaspectratio,width=\textwidth]{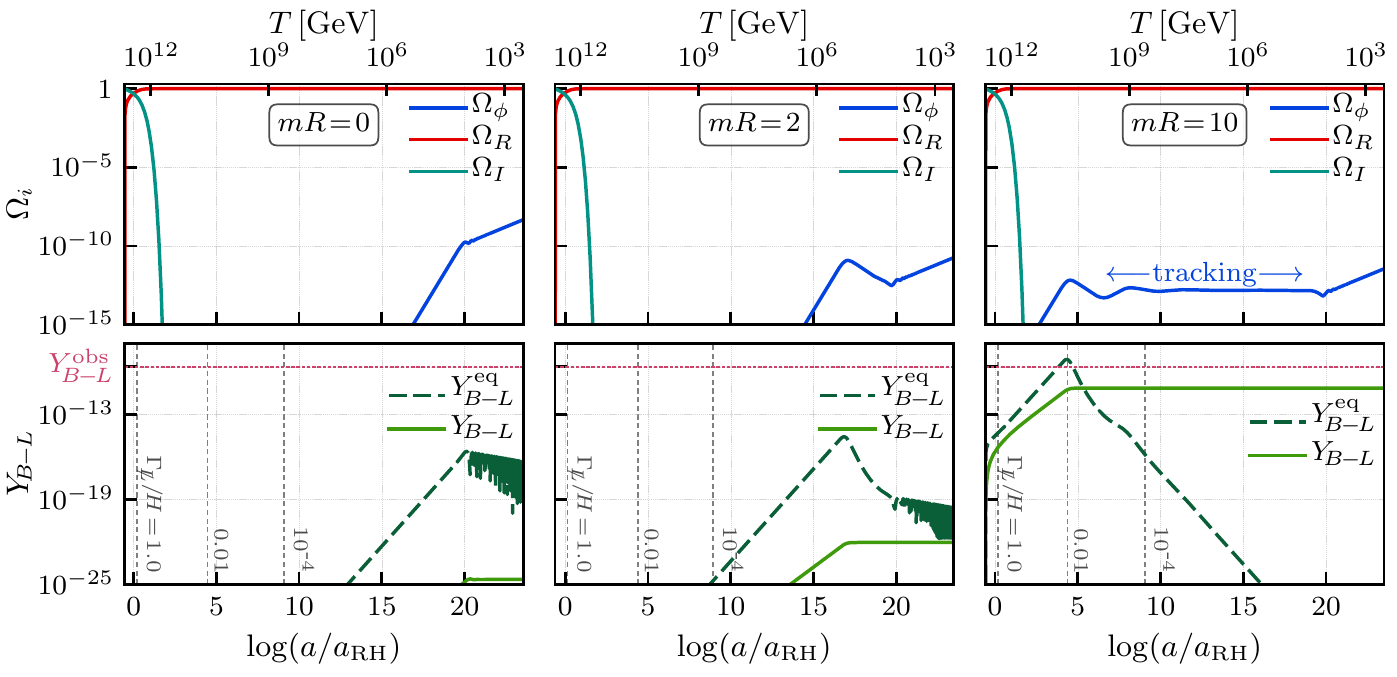}
    \end{center}
    \caption{A numerical simulation of the dynamics in the explicit realization of our scheme,
    using a continuum-clockwork axion.
    In each column, a different clockwork strength \il{mR\in [0, 2, 10]} is used.
    The rows show the evolution of cosmological abundances 
    \il{\Omega_i=\{\Omega_{I},\Omega_R,\Omegaphi\}} and 
    \il{Y_\BmL\equiv n_\BmL/s}, as a function of the number of e-folds since reheating \il{\log(a/a_{\text{RH}})}, respectively.
    The axion mass \il{m_{\phi}=1\,\eV}, initial misalignment \il{\thetain=3\pi/4}, 
    and scales for inflation \il{H_I=\Gamma_{I}=10^8\,\GeV} are fixed.  
    The equilibrium $Y_\BmL^{\rm eq}$ curves indicate that spontaneous baryogenesis 
    proceeds out-of-equilibrium in this model, as also indicated by
    the \il{\GammaLV/H} values marked by vertical gray-dashed lines.
    As $mR$ is increased, despite the mass being fixed, 
    production of asymmetry is exponentially more efficient.  In the right-hand panel 
    the ``tracking'' behavior of Sec.~\ref{subsubsec:TrackingPeriod} appears for the axion, during which the field does not oscillate, 
    but follows a radiation-like equation of state \il{\wphi\approx 1/3}, matching that of the background.
    }\label{fig:field_dynamics}
\end{figure*}

\subsection{Early dynamics and baryogenesis \label{subsec:EarlyDynamics}}

Above we have constructed the axion sector of the theory, however, we must also
incorporate the necessary ingredients for baryogenesis.  We shall proceed in
a manner parallel with Sec.~\ref{subsec:IncorporatingBaryogenesis}, specializing our
analysis to the continuum-clockwork model.  As we argued previously, the motion of the axion
\il{\p_0\phi\neq 0} spontaneously breaks $CPT$ symmetry if it
couples derivatively to a baryon current, as in Eq.~\eqref{eq:requiredterm}.  
In addition, the Weinberg operator in Eq.~\eqref{eq:wein_op} provides a source 
for processes that violate $(B-L)$. 
Assuming a similar interaction in this model,
the effective chemical potential from Eq.~\eqref{eq:muBmL} in the canonical basis
takes the form
\beq
    \muBmL(\phi,\phidot) = -g(N_f)\cdot
    \frac{\frac{1}{2\Lambda^4}\abs{\frac{\p\Veff}{\p\phi}}\dot{\phi}}{\sqrt{\left(1-\frac{\Veff}{2\Lambda^4}\right)\frac{\Veff}{2\Lambda^4}}} \ ,
\eeq
where $\Veff(\phi)$ was used for a more succinct expression.  

In order to numerically simulate the early dynamics we make several assumptions.  
Let us suppose a period of inflation with Hubble parameter \il{H_I} 
during which the axion is misaligned from the minimum of its potential by angle
\il{\thetain\in [-\pi,\pi]}.  Then, the reheating epoch is modeled by assuming the 
energy density in the inflaton \il{\rho_{I}\approx 3M_P^2H_I^2} decays into radiation $\rho_R$ 
at some rate $\Gamma_{I}$.  It follows that these quantities evolve as
\begin{align}
    \dot{\rho}_{I} + 3H\rho_{I}  &=  - \Gamma_{I}\rho_{I} \nn \\
    \dot{\rho}_{R} + 4H\rho_R &= + \Gamma_{I}\rho_{I} + \Gamma_{\phi}\rhophi \ ,
\end{align}
where the Hubble parameter is given by
\beq
H^2 = \frac{\rho_{I} + \rho_{R} + \rho_{\phi}}{3M_P^2} \ ,
\eeq
and 
\beq
\rhophi = \frac{1}{2}\dot{\phi}^2 + \Veff(\phi)
\eeq
is the energy density in the axion field.
Meanwhile, the axion equation of motion reads
\beq
\ddot{\phi} + \left(3H+\Gamma_{\phi}\right)\dot{\phi} + \pd{\Veff}{\phi} 
= \pd{\muBmL}{\phidot}\GammaLV \left(n_\BmL - n_\BmL^{\text{eq}}\right) \ .
\eeq
The term on the right-hand side is due to backreaction from $(B-L)$ generation and 
is usually negligible. As the axion is set into motion it drives the production of
$Y_\BmL$, which follows the Boltzmann evolution in Eq.~\eqref{eq:YBmLboltzmann}.

In Fig.~\ref{fig:field_dynamics} the evolution of two types of quantities --- 
the cosmological abundances \il{\Omega_i\equiv \rho_i/(3M_P^2H^2)}
and abundance $Y_\BmL$ --- are shown in the two rows of panels, 
as functions of the number of e-folds since reheating \il{\log(a/a_{\text{RH}})}.  
Almost all parameters are held fixed:
the mass \il{\mphi=1~\eV}, effective scale \il{\feff=10^{13}\,\GeV}, and 
misalignment angle \il{\thetain=3\pi/4}.  
However, we have varied
the strength of the clockwork mechanism \il{mR = \{0,2,10\}} in each 
column.  Therefore, the left-hand column shows dynamics for 
the traditional sinusoidal axion potential, the right-hand column shows a substantially 
deformed potential, and the center column shows an 
intermediate case between these two regimes.

Note that results are sensitive to inflationary scales $\Gamma_{I}$
or $H_I$ only if they are exceeded by the initial scale of 
curvature \il{\sqrt{\abs{\Veff''(\phiin)}}\approx \mphi e^{\pi mR}}.
The curvature exceeding $H_I$ implies the axion is a heavy field during inflation and 
we shall exclude this region.  On the other hand,
if the curvature exceeds $\Gamma_{I}$ it implies the axion 
is set in motion prior to reheating.  Then, washout effects that suppress the 
asymmetry can become sizeable.  In this section, 
we look to identify phenomenologically viable regions of parameter space, 
and thus as a simplifying assumption we take 
these scales to be comparable \il{\Gamma_{I}\sim H_I}.

The influence from the variation of $mR$ in Fig.~\ref{fig:field_dynamics}
is seen most immediately in the plots of $Y_\BmL$.
The deformation of the potential sets the axion in motion at higher
temperatures.  Nevertheless, we always find \il{\GammaLV/H\lesssim 1}
such that $Y_\BmL$ \emph{never exactly tracks the equilibrium value $Y_\BmL^{\rm eq}$},
\ie, this model exhibits the ``freeze-in'' spontaneous baryogenesis discussed in 
Sec.~\ref{sec:GeneralDescription}, in contrast to more common examples 
in the literature. As a result, the asymmetry is mostly set 
during the fast-roll period and remains frozen at that value.  
Using the estimate in Eq.~\eqref{eq:FIlepasy} and evaluating at the fast-roll
temperature $T_{\text{FR}}$ from Eq.~\eqref{eq:TFR} we find
\beq
\frac{Y_\BmL}{Y_\BmL^{\text{eq}}} \approx \frac{1}{3}\sqrt{\frac{1}{5}\frac{\mphi M_P}{\TD^2}}e^{\pi mR/2} \ ,
\eeq
when production ceases.  The above expression demonstrates how 
the deformation of the potential through the clockwork mechanism (\ie, the
exponential factor $e^{\pi mR/2}$) enables sufficient baryogenesis 
while maintaining a relatively light axion.

\begin{figure*}[t]
    \begin{center}
    \includegraphics[keepaspectratio,width=\textwidth]{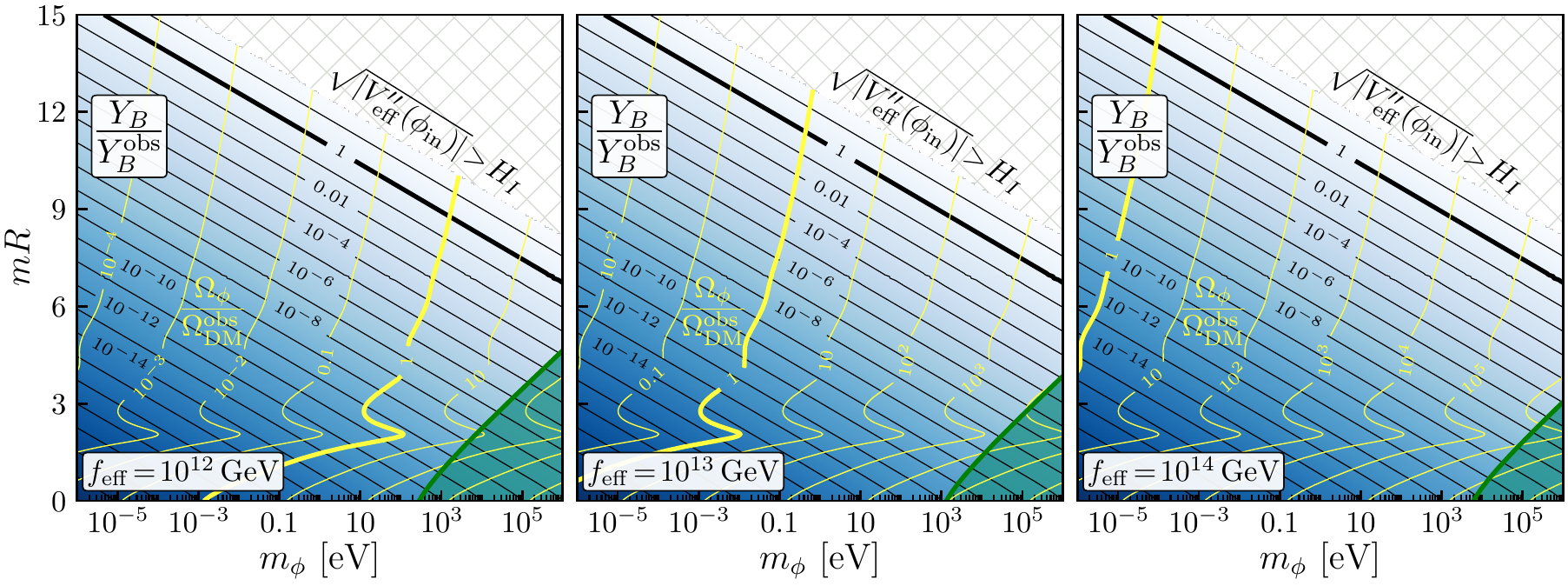}
    \end{center}
    \caption{The result of numerical simulations, showing the normalized baryon 
    abundance \il{Y_B/Y^{\text{obs}}_B} (black contours, blue shading) and axion relic abundance 
    \il{\Omegaphi/\Omega_{\text{DM}}^{\text{obs}}} (yellow contours), plotted with 
    respect to the axion mass $\mphi$ and clockwork parameter $mR$.  The two thick curves
    mark where the observed values are attained.
    The scale $\feff$ is fixed to a different value in each panel, 
    while the other parameters are held fixed to values used in previous figures: \il{H_I=10^8\,\GeV} and \il{\thetain=3\pi/4}. 
    The hatched region rules axion oscillations during inflation, while the green region shows where the 
    axion is too unstable to decays (\il{\tau_{\phi} \lesssim 10^{26}\,\text{seconds}}) 
    to serve as a DM candidate.  Other relevant constraints come from isocurvature perturbations, 
    which we cover in Sec.~\ref{sec:IsocurvaturePerturbations}.
    }\label{fig:Y_B}
\end{figure*}

Another interesting dynamical feature is found in the \il{mR=12} column
of Fig.~\ref{fig:field_dynamics}.  Although the deformation of $\Veff(\phi)$ sets the 
axion in motion earlier, it does not undergo coherent 
oscillations until much later when \il{3H\lesssim \mphi}.  
Instead, during this period the trajectory is such that $\Omegaphi$ 
is temporarily fixed, with a radiation-like equation of state \il{w_{\phi} \approx 1/3}.
Indeed, we have identified the ``tracking'' phenomenon, which we have
discussed in more generality in Sec.~\ref{subsubsec:TrackingPeriod}. 
We expect this dynamics to occur over the field range 
where \il{Z(\theta)\sim 1/\theta^2 \sim e^{\abs{\phi}/f}},
and taking \il{mR \gtrsim \mathcal{O}(\text{few})} is sufficient
to generate such a region.  We find that
\beq\label{eq:trackingpotential}
\Veff(\phi) \approx \frac{1}{2}\mphi^2f^2e^{\abs{\phi}/f}
\eeq
is a good approximation over 
\il{f \lesssim |\phi| \lesssim (\log 2 + 2\pi mR)f}.
The enhancement of the axion field range by
\il{\feff/f \approx 1 + 2mR} makes such a displacement easy to achieve.
The trajectory of the tracker is found using that
\beq\label{eq:trackingcondition}
\frac{1+\wphi}{1-\wphi} = \frac{\frac{1}{2}\phidot^2}{\Veff(\phi)} 
\eeq
is approximately constant. 
The resulting solution
\beq\label{eq:trackingsolution}
\abs{\phi(t)} \approx 
-2f\log\left[e^{-\frac{\abs{\phi_{\rm tr}}}{2f}} + \frac{\mphi \left(t-t_{\rm tr}\right)}{\sqrt{2}}\right] \ ,
\eeq
naturally depends weakly on the initial field amplitude \il{\phi_{\rm tr}\equiv \phi(t_{\rm tr})}
as it enters the tracking epoch.  Considering that tracking ends once 
\il{\mphi t\sim \mathcal{O}(1)}, the dependence on $\phi_{\rm tr}$ is
ultimately washed away if \il{\phi_{\rm tr}\gtrsim \feff}. 
Therefore, for sufficiently deformed potentials
\emph{the late-time axion field is insensitive to the initial
misalignment angle $\thetain$}, in contrast to the standard
\il{mR=0} case.

\subsection{Survey of viable regions\label{subsec:SurveyofViableRegions}}

We are now in position to discuss the phenomenological
viability of this explicit model.
As a first requirement we must verify
the existence of regions in parameter space which have both the observed dark matter
abundance \il{\Omega_{\text{DM}}^{\text{obs}} \approx 0.26} and baryon abundance
\il{Y_B^{\text{obs}} \approx 8.6\times 10^{-11}}.  
Furthermore, regions in which the axion is not sufficiently 
stable, \ie, lifetimes longer 
than \il{\tau_{\phi} \gtrsim 10^{26}\,\sec}~\cite{Chen:2003gz,Zhang:2007zzh},
must be excluded.  Indeed, regions with substantial deformations, \ie, at least moderately 
large $mR$, are where we expect to find viability in these respects.

As we have found above, such a regime is also associated with a tracking period
for the axion.  While tracking has little direct effect on 
the development of baryon asymmetry, it does have an marked 
influence on the relic abundance $\Omegaphi$.
Namely, employing the tracking-field solution in Eq.~\eqref{eq:trackingsolution} 
we find at the onset of coherent oscillations
\il{\rhophi\approx \mphi^2f^2}, so that at present day
\beq\label{eq:abundance}
\Omegaphi h^2 \approx 
0.12 \left(\frac{\feff}{10^{13}\,\GeV}\cdot \frac{12}{mR}\right)^2 \sqrt{\frac{\mphi}{0.53~\eV}}  \ ,
\eeq
In the region of interest \il{mR\gtrsim\mathcal{O}(\text{few})} 
we find agreement with numerical computations to \il{\lesssim 10\%}.
Note that the insensitivity of $\Omegaphi$ to the initial misalignment angle 
$\thetain$ is a result of the attractor-like dynamics and 
distinguishes our result from the standard axion cosmology.

An analytical approximation for the baryon asymmetry can also be constructed
using the general result in Eq.~\eqref{eq:FIlepasy},
taking the integrated $\mathcal{O}(1)$ factor to be unity.
We find an approximate expression 
\beq
Y_B \sim 10^{-10}\left(\frac{\mphi}{0.53~\eV}\right) e^{\pi (mR-12)} \ ,
\eeq
which holds to at least order-of-magnitude accuracy throughout the parameter space.

The results of our numerical simulations span the \il{\{\mphi,mR,\feff\}} space 
and are shown in Fig.~\ref{fig:Y_B}.  In each panel, contours show both 
the normalized baryon abundance \il{Y_B/Y_B^{\text{obs}}} (black) and 
axion abundance \il{\Omegaphi/\Omega_{\text{DM}}^{\text{obs}}} (yellow).  
The sole distinction between each panel is the choice for the scale $\feff$.
The green regions show exclusions due to axion decays.  As expected from
Eq.~\eqref{eq:decaysuppression}, for even moderately large $mR$ these regions 
are substantially reduced in size. 
The points of intersection between the two thickest curves correspond to viable 
configurations that match observations, and these plots show that \emph{viable regions
exist over all the panels shown} in Fig.~\ref{fig:Y_B}.
The only constraints not yet applied are bounds on isocurvature perturbations,
which is the focus for the remainder of the section.

\subsection{Isocurvature perturbations\label{sec:IsocurvaturePerturbations}}

In the more general analysis provided in Sec.~\ref{subsec:IsocurvaturePerturbations},
several significant observations were made regarding the axionic and baryonic 
isocurvature perturbations.  Moreover, power spectra for these perturbations were found 
for both in-equilibrium and out-of-equilibrium production of the baryon asymmetry.  
We conclude this section with a more thorough treatment, in which the perturbation
equations are solved numerically, within the context of the continuum-clockwork axion model.

\subsubsection{System of perturbations\label{subsubsec:SystemofPerturbations}}

In our analysis, for perturbations of the FRW background, 
we use the conformal Newtonian gauge, 
defined by the line element~\cite{Bardeen:1980kt,Mukhanov:1990me}:
\beq
ds^2 = (1+2\Phi)dt^2 - a(t)^2(1 - 2\Psi)d\vec{x}^2 \ ,
\eeq
where the scalar potentials \il{\Phi,\Psi} are functions of space and time.  
The anisotropic stress is vanishing in our model, which implies 
an equivalence \il{\Psi = \Phi}.  

The gravitational potential develops
according to the Einstein equations as~\cite{rubakovinflation}
\beq
    \frac{k^2}{a^2}\Phi + 3H\left(H\Phi + \dot{\Phi}\right)  = -\frac{\delta\rho_{\text{tot}}}{2M_P^2}
\eeq
for a comoving Fourier mode $k$, where 
\il{\delta\rho_{\text{tot}}} is the sum of energy density perturbations.  
Meanwhile, covariant stress-energy conservation gives the
evolution for matter degrees of freedom. Namely, it gives axion perturbations
\beq
\delta\ddot{\phi} + 3H\delta\phidot + \left[\frac{k^2}{a^2} + \Veff''(\phi)\right]\delta\phi = 4\phidot\dot{\Phi} - 2\Veff'(\phi)\Phi \ ,
\eeq
and radiation perturbations
\beq
\dot{\delta}_{\gamma} - \frac{4}{3} k^2 v_{B\gamma} = 4\dot{\Phi} \ ,
\eeq
as well as the velocity potential $v_{B\gamma}$ of the baryon-photon fluid.  However, 
for large-scale perturbations in our scenario $v_{B\gamma}$ has a negligible influence.
Finally, the perturbations in lepton or baryon density are
coupled to the axionic degrees of freedom through \il{\delta\muBmL=\delta\dot{\theta}(\dphi,\dphidot)}:
\beq\label{eq:dnLeqn}
\dot{\delta}_B - \frac{k^2}{a^2}v_{B\gamma} 
= -\GammaLV\left(\delta_B - \frac{\delta \muBmL}{\muBmL}\right)\frac{Y_B^{\text{eq}}}{Y_B} + 3\dot{\Phi} \ ,
\eeq
where the distinction between $Y_B$ and $Y_\BmL$ as they appear in these equations 
is inconsequential.

\subsubsection{Initial conditions}

Before discussing the features of this system in some detail,
let us first make our initial conditions and other ancillary assumptions clear.  
The isocurvature mode is formally defined by a vanishing initial condition for the gauge-invariant 
curvature perturbation~\cite{Mukhanov:1990me}
\beq\label{eq:Rperturbation}
\mathcal{R} \equiv \frac{2}{3}\frac{H\Phi + \dot{\Phi}}{\left(1+w\right)H} + \Phi \ ,
\eeq
and it follows that $\Phi$, $\dot{\Phi}$, and 
$\delta\rho_{\text{tot}}$ (after enforcing the Einstein equations)
all have vanishing initial conditions as well~\cite{Perrotta:1998vf}.

The stress-energy fluctuations are functions
both of perturbations in the field and the gravitational potential:
\begin{align}
    \delta\rhophi &= \phidot\delta\dot{\phi} - \phidot^2\Phi + \Veff'(\phi)\dphi \nn \\
    \delta P_{\phi} &= \phidot\delta\dot{\phi} - \phidot^2\Phi - \Veff'(\phi)\dphi \ ,
\end{align}
so there is a non-zero initial axion perturbation
\beq\label{eq:initialaxionperturbation}
\delta_{\phi} \simeq \frac{\Veff'(\phiin)}{\Veff(\phiin)}\frac{H_I}{2\pi} \ .
\eeq

Note that in a more traditional scenario --- \eg, the QCD axion ---
the potential is flat until the confining phase transition is 
approached, when it is finally generated by instanton effects.   
Although the same non-zero field fluctuation $\delta\phi$ exists,
the perturbation $\delta_{\phi}$ is vanishing until the potential
for the axion is generated.  
By contrast, in our case $\Veff(\phi)$ is established
already during (or prior to) inflation.  As a result of this contrast
and the deformation of the potential in our model, we shall find 
several interesting features in the evolution of axionic perturbations,
even for large-scale modes.

\begin{figure}[t]
    \begin{center}
    \includegraphics[keepaspectratio,width=0.49\textwidth]{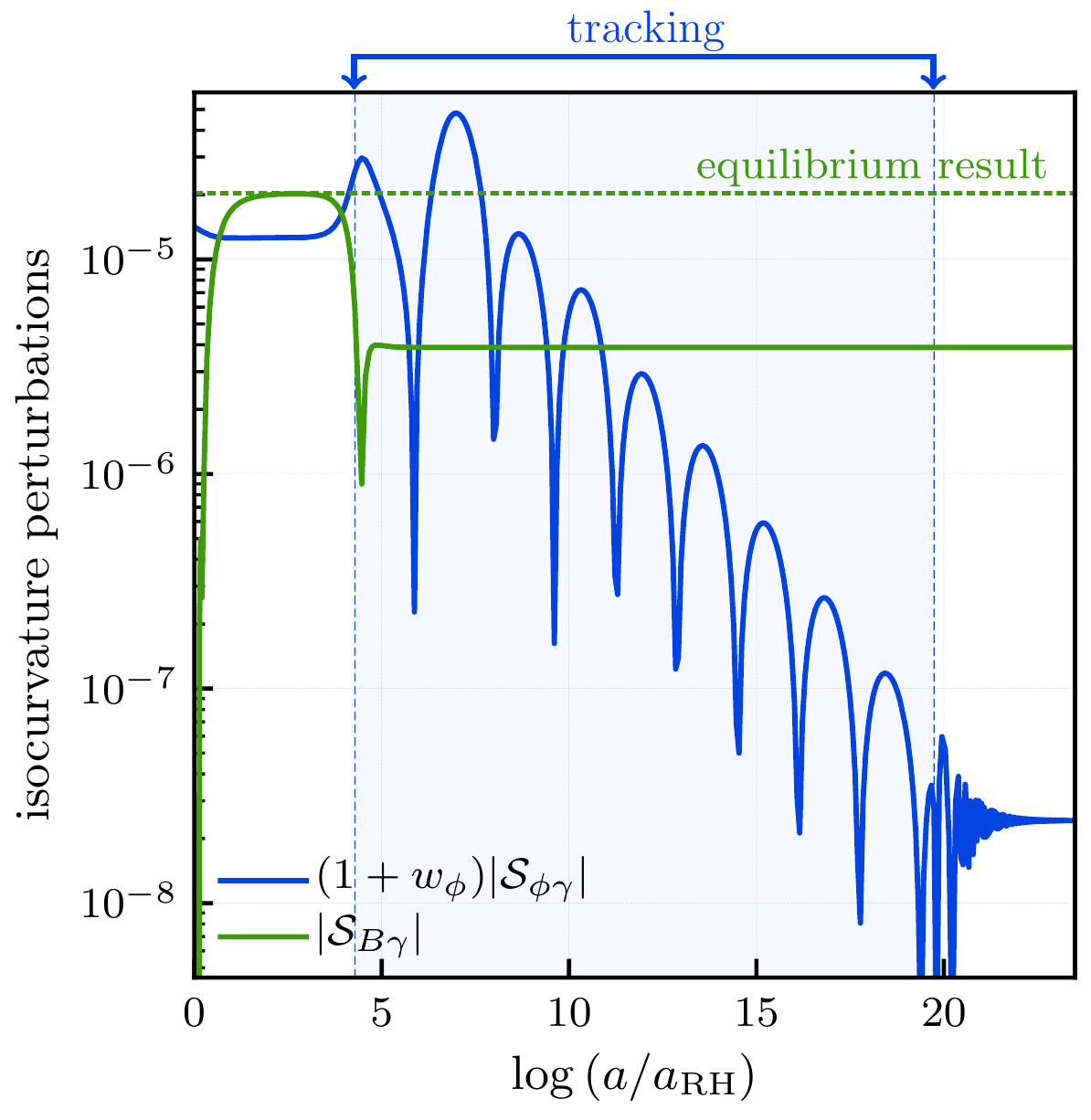}
    \end{center}
    \caption{The evolution of both the axionic $\mathcal{S}_{\phi\gamma}$ (blue curve) 
    and baryonic $\mathcal{S}_{B\gamma}$ (green curve) contributions to the isocurvature mode, 
    with a choice of parameters --- \il{\mphi=1~\eV}, \il{mR = 10}, \il{\feff = 10^{13}\,\GeV}, 
    \il{\thetain = 3/4}, \il{H_I=10^8\,\GeV} --- similar to previous figures.  
    The dashed-green curve shows the result in Eq.~\eqref{eq:equilibriumSBgamma},
    if spontaneous baryogenesis had occurred at equilibrium, and the shaded-blue 
    area indicates the tracking phase, suppressing as \il{\mathcal{S}_{\phi\gamma}\propto 1/\sqrt{a}} over its duration.
    }\label{fig:perturbation_dynamics}
\end{figure}

\subsubsection{Axionic contribution}

There are several significant observations to make that are unique to the
axionic contribution.  To simplify the discussion, we momentarily
ignore the baryonic component and define two gauge-invariant entropy perturbations. One is 
intrinsic~\cite{Kodama:1985bj}:
\beq\label{eq:Gammaperturbation}
\Gamma \equiv \frac{\dPphi/\rhophi - c_{\phi}^2\delta_{\phi}}{1-c_{\phi}^2} \ ,
\eeq
where \il{c_{\phi}^2\equiv \dot{P}_{\phi}/\dot{\rho}_{\phi}} is the adiabatic 
sound speed of the axion fluid.
The other is expressed relative to photons:
\beq\label{eq:SphiRperturbation}
\mathcal{S}_{\phi\gamma} \equiv  \frac{\delta_{\phi}}{1+\wphi} - \frac{3}{4}\delta_{\gamma} \ .
\eeq
The relevant modes for our discussion are outside the
Hubble sphere \il{k \ll aH} during
the early evolution.  At these scales, the basis of 
gauge-invariant perturbations \il{\{\Phi,\mathcal{R}\}} and 
\il{\{\Gamma, \mathcal{S}_{\phi\gamma}\}} 
is convenient, since these two sets decouple.  
In particular, writing the perturbation equations in this basis
and simplifying the system for the tracking regime (\ie, taking \il{\wphi\rightarrow w=\frac{1}{3}}), 
we find~\cite{Bartolo:2003ad}
\begin{align}\label{eq:trackingperturbationeqns}
    \frac{1}{2}\frac{d\left[(1+\wphi)\mathcal{S}_{\phi\gamma}\right]}{d\log a} &= -\Gamma \nn \\
    2\left[(1+\wphi)\mathcal{S}_{\phi\gamma}\right] - \Gamma &= \frac{d\Gamma}{d\log a} \ .
\end{align}
The solutions for $\mathcal{S}_{\phi\gamma}$ undergo damped harmonic oscillations
every few e-folds, rapidly suppressing the 
axionic isocurvature amplitude as\footnote{Similar effects have appeared in the literature on
investigations of quintessence field perturbations~\cite{Abramo:2001mv,Kawasaki:2001nx,Copeland:2006wr}.}
\beq
\mathcal{S}_{\phi\gamma} \propto \frac{1}{\sqrt{a}} \ .
\eeq
It is instructive to view a numerical solution of the full 
system of perturbations in this regime, focusing on the tracking period.  
In Fig.~\ref{fig:perturbation_dynamics} this is shown by the blue curve, where we have 
highlighted the tracking phase.  Indeed, a few e-folds after reheating the
axion begins to converge to the tracker solution, and the amplitude of the isocurvature perturbation 
falls as \il{1/\sqrt{a}}.  The field eventually enters a region of the potential
which is approximately harmonic, ending the tracking dynamics and 
thus concluding the suppression.
The isocurvature $\mathcal{S}_{\phi\gamma}$ then
undergoes some short-lived transient oscillations before 
finally settling on its asymptotic late-time value.

\subsubsection{Baryonic contribution}

The dynamical suppression of $\mathcal{S}_{\phi\gamma}$ is significant, 
since it could ensure that the baryonic contribution is dominant
if tracking lasts for a sufficient number of e-folds.  Indeed, this dominance was an
assumption we made when deriving the general power 
spectra in Sec.~\ref{subsec:IsocurvaturePerturbations}.
The baryonic component is defined in an analogous way:
\beq
\mathcal{S}_{B\gamma} \equiv \frac{\delta Y_B}{Y_B} = \delta_B - \frac{3}{4}\delta_{\gamma} \ .
\eeq 

An example of numerical solutions for the evolution of $\mathcal{S}_{B\gamma}$
is shown in the solid-green curve of Fig.~\ref{fig:perturbation_dynamics}.
Apart from some early dynamical behavior as the baryon asymmetry is being established,
the $\mathcal{S}_{B\gamma}$ perturbations are essentially fixed after the fast-roll period.
However, there are other subtleties that we should outline.

Although the baryon asymmetry is typically produced out-of-equilibrium in 
this model, a useful benchmark comparison with regard to the perturbation spectrum is 
the case of in-equilibrium production, which we have shown 
as a dashed-green line in Fig.~\ref{fig:perturbation_dynamics}.  
We find that the baryonic perturbations for different types of production 
typically do not differ by more than an order of magnitude throughout most of parameter space.  
However, there are some particularly important exceptions.
To this end, it is instructive to derive an analytical approximation for the equilibrium
case in terms of the canonical field $\phi$.  
Using the proportionality \il{Y_\BmL\approx Y_\BmL^{\text{eq}} \propto \muBmL T^2}
we find the perturbation by evaluating
\beq\label{eq:equilibriumSBgamma}
\delta_B \approx \frac{\delta \muBmL}{\muBmL} = 
\frac{1}{\muBmL}\left(\frac{\p\muBmL}{\p\phi}\dphi + \frac{\p\muBmL}{\p \dot{\phi}}\delta \dot{\phi}\right)
\eeq
at the decoupling temperature $\TD$. 
In previous investigations (\eg, Ref.~\cite{DeSimone:2016ofp}), a slow-roll
approximation is used and neither $\delta\phi$ or $\phi$ are assumed to 
significantly evolve.  Under these conditions, we find an analytical expression:
\beq\label{eq:equilibriumSBgamma2}
S_{B\gamma} \approx 
\left.\left\{\frac{2\Veff''(\phi)}{\Veff'(\phi)}
-\frac{\Veff'(\phi)}{\Veff(\phi)}\left[\frac{1-\frac{\Veff}{\Lambda^4}}{1-\frac{\Veff}{2\Lambda^4}}\right]\right\}\right|_{\phiin}\!\!\frac{H_I}{2\pi} \ ,
\eeq
where the form of the expression is influenced by the chemical potential 
being a function of both the velocity of the canonical field and 
the field itself \il{\muBmL\propto \dot{\theta}(\phi,\phidot)}.
Additionally, note the appearance in Eq.~\eqref{eq:equilibriumSBgamma2} 
of several critical points for the initial 
field displacement $\phiin$: the two terms may 
have opposite signs, allowing for cancellations and a vanishing $\mathcal{S}_{B\gamma}$,
and any inflection points in the canonical potential cause the first term to vanish.

The behavior of the perturbations near these points can dramatically change the isocurvature
power spectrum, making the baryonic contribution subdominant.
While the perturbations for out-of-equilibrium asymmetry production cannot be 
found analytically in this way, it is important to investigate how these
effects are manifested in that case, which is a question we shall continue to address below.

\begin{figure}[t]
    \begin{center}
    \includegraphics[keepaspectratio,width=0.49\textwidth]{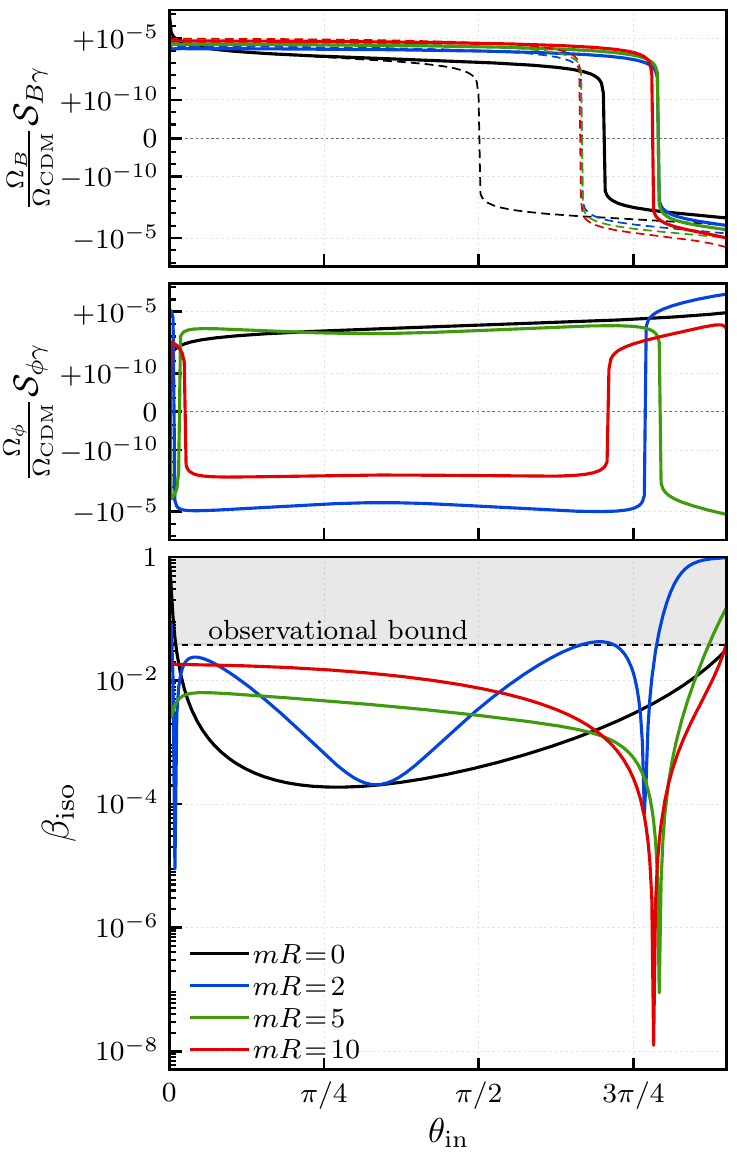}
    \end{center}
    \caption{In the top two panels, numerical results for the baryonic and 
    axionic contributions to the isocurvature power $\mathcal{P}_{\mathcal{S}\mathcal{S}}(k_*)$ 
    are shown, respectively, as a function of the misalignment angle $\thetain$,
    with the various curves showing different values of $mR$.  The values
    \il{\feff=10^{13}\,\GeV} and \il{H_I=10^8\,\GeV} were chosen, and 
    $\mphi$ is set to satisfy \il{\Omegaphi\approx\Omega_{\text{DM}}^{\text{obs}}}.
    In the top panel, dashed curves show the approximate equilibrium result 
    of Eq.~\eqref{eq:equilibriumSBgamma2}.  In the bottom panel, the isocurvature 
    fraction $\beta_{\text{iso}}$ is computed for each curve, comprising the 
    total effect.
    }\label{fig:iso_contours}
\end{figure}

\begin{figure*}[t]
    \begin{center}
    \includegraphics[keepaspectratio,width=\textwidth]{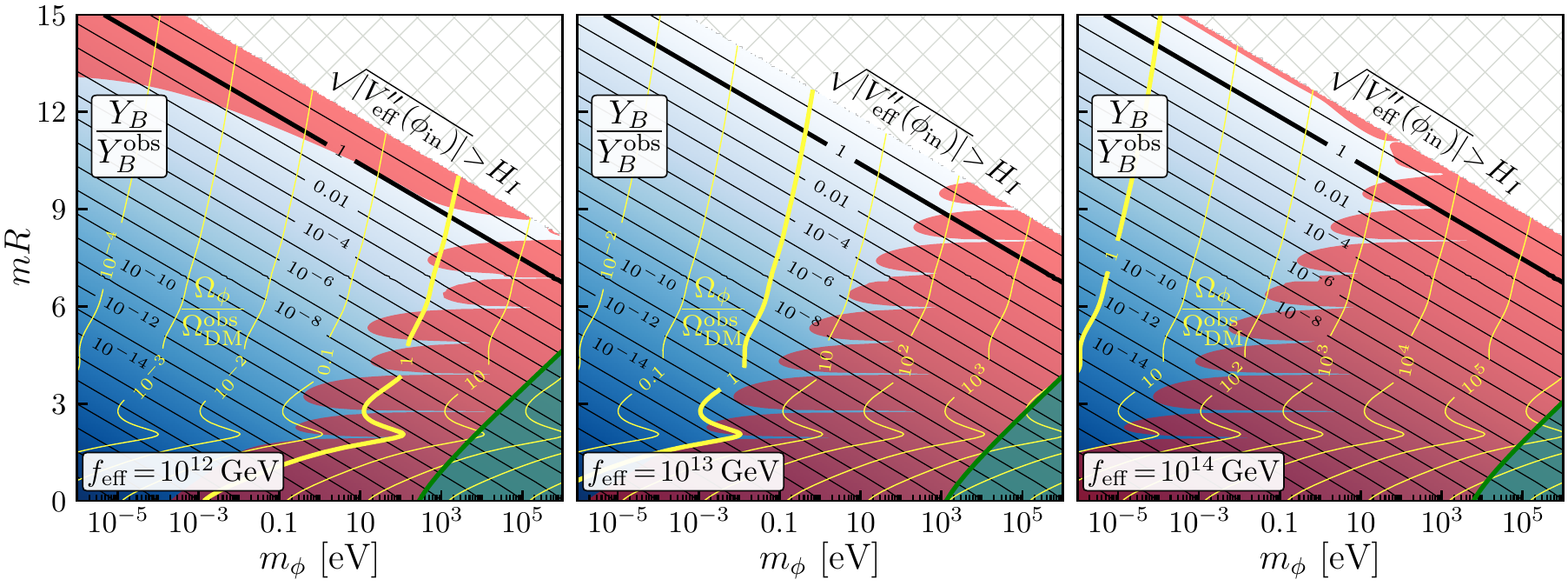}
    \end{center}
    \caption{The result from Fig.~\ref{fig:Y_B}, now including exclusion regions from isocurvature constraints (in red).
    }\label{fig:Y_B_iso}
\end{figure*}

\subsubsection{Isocurvature bounds}

The constraint on isocurvature from the CMB comes in the form of an 
upper-bound on the uncorrelated ``isocurvature fraction,''
from \emph{Planck} collaboration data~\cite{Akrami:2018odb}:
\beq\label{eq:betaiso}
\beta_{\text{iso}}(k_*) \equiv \frac{\mathcal{P}_{\mathcal{SS}}(k_*)}{\mathcal{P}_{\mathcal{RR}}(k_*) + \mathcal{P}_{\mathcal{SS}}(k_*)} < 0.038 \ ,
\eeq
in which \il{\mathcal{P}_{\mathcal{RR}}(k_*)\approx 2.10\times 10^{-9}} is the adiabatic 
power, $\mathcal{P}_{\mathcal{SS}}(k_*)$ is the isocurvature power, and each is evaluated at 
the pivot scale \il{k_*\equiv 0.05~\Mpc}.  The baryonic and axionic contributions 
are exactly correlated due to their common source,  and appear as a weighted sum
\beq\label{eq:PSS}
\mathcal{P}_{\mathcal{SS}}(k) \equiv 
\left[\frac{\Omegaphi}{\OmegaCDM}\mathcal{S}_{\phi\gamma}(k) + \frac{\Omega_B}{\OmegaCDM}\mathcal{S}_{B\gamma}(k)\right]^2 \ ,
\eeq
where \il{\Omega_B\approx 0.0486} and \il{\OmegaCDM\approx 0.2589}
are the cosmological abundances for baryons and cold dark matter (CDM), respectively.

Note that the total power spectrum $\mathcal{P}_{\mathcal{S}\mathcal{S}}$
has the possibility for cancellations \emph{between} axionic and baryonic components.
This type of behavior is made more clear in the context of our model by 
examining the perturbations as a function of the misalignment angle $\thetain$.
In the bottom panel of Fig.~\ref{fig:iso_contours}, 
we plot the dependence of $\beta_{\text{iso}}(k_*)$ 
on $\thetain$ by numerically solving the perturbation equations.  
The upper panels show explicitly how the 
weighted isocurvature sources in Eq.~\eqref{eq:PSS} contribute to the bottom
panel.  The different curves show various choices for $mR$, while $\mphi$ is taken
to ensure the axions have the observed DM abundance \il{\Omegaphi\approx \OmegaCDM}
at \il{\thetain=3\pi/4} (the value used in all previous figures). 
We have also included the equilibrium result from Eq.~\eqref{eq:equilibriumSBgamma2}
with dashed curves in the top panel.  Our interest is mostly in the behavior for
at least moderate values of $\thetain$, to ensure the field is misaligned sufficiently
from the minimum to enable adequate production of baryon asymmetry.

In the \il{mR=0} case of Fig.~\ref{fig:iso_contours}, 
the axionic isocurvature is monotonic with $\thetain$
and does not experience any sign changes.
However, as $mR$ is increased  both isocurvature contributions 
show vanishing points that generally do not coincide. 
We also confirm that as $mR$ is increased tracking effects 
suppress the axionic component, as seen through the overall 
reduction in the $\mathcal{S}_{\phi\gamma}$ amplitude.
The effect is more subtle toward the 
edge of field space, however, as both contributions are enhanced with $\thetain$.
The accumulation of all the effects is that as we deform the potential,
the baryon asymmetry is amplified exponentially as $e^{\pi mR}$, 
while the isocurvature is \emph{increasingly suppressed} at moderately
large misalignment angles, focused roughly around the \il{\thetain\sim 3\pi/4} region. 

While these discussions are instructive in forming a qualitative picture of
the perturbations, our interest ultimately is in producing exclusion regions over
the plots in Fig.~\ref{fig:Y_B}.
Therefore, we solve the perturbation equations over the full parameter space
and mark regions that violate Eq.~\eqref{eq:betaiso}.  These are indicated by dark-red 
in Fig.~\ref{fig:Y_B_iso}, while all other features and parameter choices are identical
to Fig.~\ref{fig:Y_B}, as discussed previously.  
We immediately observe that the viable regions in which the 
baryon abundance $Y_B$ (black contours) and axion abundance $\Omegaphi$ (yellow contours)
are produced in the observed amounts \emph{remain safely outside the exclusion region}.

\section{Discussion and Conclusions\label{sec:Conclusions}}


In this paper, we have investigated the possibility that both the 
baryon asymmetry of the universe and dark matter may 
be accounted for by a single axion-like field.  
In this scenario, the early-universe dynamics of the axion 
drive a period of spontaneous baryogenesis, during which the 
observed baryon asymmetry is produced.  As the axion field settles 
to the minimum of its potential, it undergoes coherent oscillations, 
which behave cosmologically as dark matter at late times.  
Typically, to generate the observed baryon asymmetry, a relatively 
``steep'' axion potential is required in the region where the axion 
initially rolls.  The corresponding axion mass is large and 
highly unstable against decays, making it inadequate as a dark matter candidate.  
However, we have shown that a field-dependent 
wavefunction renormalization can arise which effectively ``deforms'' 
the axion potential, inducing in a mismatch in curvature between different
regions.  In this way, novel possibilities have emerged, 
as we can not only generate the observed baryon and dark 
matter abundance jointly, but the axion dynamics 
can also exhibit dramatic modifications.

In Sec.~\ref{sec:GeneralDescription}, we have given a general 
description of the type of wavefunction renormalization
necessary to realize such a scenario.  Namely, with an enhancement 
\il{Z(\theta\lesssim \epsilon)\simeq 1/\epsilon^{2n}} 
near the minimum of the axion potential, and 
\il{Z(\theta\simeq\thetain)=\mathcal{O}(1)} near the edges, 
the necessary deformations in the canonical potential are generated.  
Specifically, for \il{\epsilon \ll 1} this has 
the effect of flattening the potential near its minimum
while leaving its shape toward the edges of field space unaltered.  
The late-time mass is then suppressed by a factor of \il{\epsilon^n}, 
while the effective chemical potential which efficiently drives
spontaneous baryogenesis is retained.  Moreover, the wavefunction
enhancement also has the effect of suppressing the axion decay width
by a factor of \il{\epsilon^{2n}}.
As we have discussed, the culmination of these features is that the 
general arrangement in Sec.~\ref{sec:GeneralDescription} can yield
the observed baryon asymmetry, while maintaining a sufficiently light and
stable axion dark matter candidate.  We have investigated the production of
baryon asymmetry --- both in-equilibrium and out-of-equilibrium ---
and found that both cases present compelling possibilities.

Meanwhile, to interpolate between the two regions of the wavefunction, 
we implemented a simple power-law form \il{Z(\theta) \simeq 1/\theta^{2n}}.  
As a result, we have shown that the axion exhibits a ``tracking'' behavior as it transits 
through this region. The field follows an attractor-like trajectory in which its late-time
evolution is made increasingly insensitive to initial conditions. 
This phenomenon implies not only an axion relic abundance which is insensitive 
to the initial misalignment angle, but also a suppression of its isocurvature perturbations.
We also have described how the axion equation-of-state parameter
during this period converges to a non-trivial 
value \il{\wphi\rightarrow (1+w-n)/n}, which reflects the shape of the potential 
through its dependence on the parameter $n$.  

In Sec.~\ref{sec:AnExplicitModel}, we have supplied a ``proof of concept''
by constructing an explicit model using the five-dimensional 
continuum-clockwork axion, which serves as a realization of the
more general scenario described in Sec.~\ref{sec:GeneralDescription}.  In particular, by 
integrating out the heavy KK modes and examining the theory for the lightest 
four-dimensional axion, we have shown that such a model furnishes 
a wavefunction renormalization $Z(\theta)$ with similar properties 
to the \il{n=1} case of Sec.~\ref{sec:GeneralDescription}.
The small parameter $\epsilon$ that determines the deformation of the axion
potential is mapped onto a factor $e^{\pi mR}$ in the clockwork theory, such that the
scale of bulk and boundary masses $m$ and the size of the extra dimension $R$
together set the strength of the deformation.
We have shown (see, for example, Fig.~\ref{fig:field_dynamics})
that spontaneous baryogenesis in this model is typically accomplished
via out-of-equilibrium production, in contrast to many of the conventional 
spontaneous baryogenesis models in the literature.  Moreover, we have also recovered the anticipated
tracking dynamics in this model, as the clockwork parameters 
exceed \il{mR\gtrsim \mathcal{O}(\text{few})}.  
We have determined regions of phenomenological viability by 
producing a set of numerical simulations over the parameter space.  
Namely, in Fig.~\ref{fig:Y_B} we have shown the produced baryon asymmetry and 
axion relic abundance over a range of parameters and found several viable regions.

We have also, at the close of Sec.~\ref{sec:AnExplicitModel}, given a more
thorough treatment of the large-scale isocurvature perturbations 
produced in this model, which include both axionic and baryonic components. 
The evolution of these components in the early universe
is made non-trivial by the deformations to the potential.  As 
anticipated in Sec.~\ref{sec:GeneralDescription}, the axionic component is suppressed
by tracking dynamics, and we have determined that in most regions
the baryonic isocurvature component is dominant.
Furthermore, we have demonstrated an interesting dependence of the perturbations
on the initial misalignment angle $\thetain$.  There are certain critical points 
for $\thetain$ where sign-changes can occur in the amplitude of 
either of the perturbation components, which can result in a suppression in that region.
These points generally shift throughout the model parameter
space.  The culmination of these effects is a non-trivial bound imposed by the CMB
isocurvature constraints.  Although these bounds can be quite severe, 
we have found that the viable regions for the CCW model
all remain below the isocurvature constraints (see Fig.~\ref{fig:Y_B_iso}).

To conclude, we have shown in this paper that an axion with a field-dependent 
wavefunction renormalization, which is enhanced near the minimum of the axion potential, 
can generate both the observed baryon asymmetry and dark matter relic abundance.  
Using the continuum-clockwork axion, we have constructed an explicit model 
realization of this idea.  Our results also suggest directions
for further research,  including approaches with multiple scalar fields,
where non-trivial dynamics can arise that significantly alters the effective chemical
potential, \eg, effects from temperature-dependent masses~\cite{Dienes:2015bka,Dienes:2016zfr}.
Moreover, the CCW axion model constitutes only a single realization 
of the more general idea in this paper.  A natural extension is to 
explore other models which yield similar non-canonical 
kinetic terms, but an altogether different set of phenomenological possibilities.

\bigskip

\FloatBarrier
\section*{Acknowledgments\label{sec:Acknowledgments}}


The authors thank Natsumi Nagata and Lauren Pearce for useful discussions.
This work was supported by IBS under the project code IBS-R018-D1.

\appendix

\FloatBarrier
\section{Tracking Dynamics\label{sec:TrackingDynamics}}


In this appendix, we provide a short review on cosmological tracking solutions 
and their general classification in terms of the scalar field potential.
Throughout much of this review we closely follow the methodology and 
results of Ref.~\cite{Steinhardt:1999nw}.
We first give the necessary background for Sec.~\ref{sec:GeneralDescription}
and the formalism used to deduce the class of potentials and regions
in field space which exhibit tracking solutions [see Eq.~\eqref{eq:tr_pot}].
Then, we tailor our analysis specifically to the continuum-clockwork 
axion example of Sec~\ref{sec:AnExplicitModel}, showing that tracking 
solutions are a generic property of these potentials which drive the axion 
equation of state to that of the background \il{\wphi\rightarrow w}.

\subsection{Classification of Tracking Potentials\label{subsec:Classification}}

A tracking field, by definition, is a field that converges to a given evolution
in phase space, even under a variation in initial conditions.
Typically, such attractor-like solutions are also associated with convergence of the 
equation-of-state parameter $\wphi$ to some fixed value, but this 
ultimately depends on the background cosmology.

Let us consider a canonically normalized scalar field $\phi$ 
with potential $V(\phi)$ that evolves in an FRW spacetime:
\beq\label{eq:appendixeqnofmotion}
\ddot{\phi} + 3H\dot{\phi} + V'(\phi) = 0 \ .
\eeq
A useful parameter to define is the ratio of
kinetic energy to potential energy of the scalar field:
\beq\label{eq:xdef}
x \equiv \frac{\frac{1}{2}\dot{\phi}^2}{V(\phi)} = \frac{1+\wphi}{1-\wphi} \ .
\eeq
After some rearrangement, Eq.~\eqref{eq:appendixeqnofmotion} can be recast as 
\beq\label{eq:appendixeqnofmotion2}
\frac{1}{6}\frac{d\log x}{d\log a} = M_P\sqrt{\frac{\Omegaphi}{3(1+\wphi)}}\abs{\frac{V'(\phi)}{V(\phi)}} - 1 \ .
\eeq
A tracking solution with a convergent equation of state requires that 
$x$ is approximately constant, such that the last term is small.  The expression
\beq\label{eq:trackingcond}
\abs{\frac{V'(\phi)}{V(\phi)}} \approx \frac{1}{M_P}\sqrt{\frac{3(1+\wphi)}{\Omegaphi}}
\approx \frac{H}{\big|\dot{\phi}\big|}
\eeq
then dictates the tracking trajectory, where in the last approximation 
we implicitly assumed \il{1+\wphi=\mathcal{O}(1)}.  

Naturally, for the tracking condition in Eq.~\eqref{eq:trackingcond} to remain 
satisfied as the system evolves, both sides of the relation must change in the same way.
Therefore, differentiating the equation of motion with respect to $\phi$, and demanding
still that $x$ varies negligibly with time, we arrive at the relation
\beq\label{eq:Gammadef}
\Gamma \equiv \frac{V''(\phi)V(\phi)}{\left[V'(\phi)\right]^2} \approx 1 + \frac{w-\wphi}{2(1+\wphi)} \ .
\eeq
The necessary (but not sufficient) condition is that a
region of the potential may yield tracking solutions if the
dimensionless quantity $\Gamma$ does not vary appreciably over that field range.
It acts as a determinant for classifying the features of different tracking regions
and it does this \emph{only through properties of the potential},
without reference to any dynamical information.  
In particular, the equation-of-state parameter to which the tracker converges is found
by rearranging the above expression:
\beq\label{eq:trackingw}
\wphi \approx \frac{w - 2(\Gamma-1)}{1+2(\Gamma-1)} \ ,
\eeq
which is determines the evolution of $\Omegaphi$ during tracking.

The condition in Eq.~\eqref{eq:Gammadef} is not always sufficient
because it does not guarantee that the tracking solutions
are stable under small perturbations to the equation of state.
An analysis shows that
\beq
\Gamma > 1 - \frac{1 - w}{6+2w} \geq \frac{1}{2}
\eeq
is required for stable tracking solutions.\footnote{More specifically, for 
a matter-dominated epoch this implies \il{\Gamma > \frac{5}{6}} and 
for a radiation-dominated epoch \il{\Gamma > \frac{9}{10}}.}
Moreover, within the above range there are two distinctive behaviors.
In the case that \il{\Gamma > 1}, the equation of state for the scalar field 
is less kinetic than the background \il{\wphi < w}, so the
abundance $\Omegaphi$ grows during tracking.  
On the other hand, in the \il{\Gamma < 1} case we find \il{\wphi > w} 
instead, and the abundance falls during that epoch.
The ``borderline'' scenario of \il{\Gamma=1} is also an interesting critical case for
which $\wphi$ is driven to match the background, and $\Omegaphi$
does not evolve at all.  This borderline case is found with potentials 
that have an exponential region.  
Incidentally, this is approximately the scenario we find in the continuum-clockwork axion example 
of Sec.~\ref{sec:AnExplicitModel}, for which we now briefly specialize our discussion.

\subsection{Tracking with Continuum-Clockwork Axion\label{subsec:CCWTracking}}

Let us now examine the continuum-clockwork example of Sec.~\ref{sec:AnExplicitModel}
and use the analysis above the identify any tracking regions for that potential.
It is perhaps instructive to first consider \il{mR=0}, \ie, the
standard sinusoidal axion potential \il{\Veff(\phi)=\Lambda^4[1-\cos(\phi/f)]}.
Using the definition in Eq.~\eqref{eq:Gammadef} we find that 
\beq
\Gamma = 1 - \frac{1}{1 + \cos\left(\frac{\phi}{f}\right)} \ .
\eeq
Regardless of how slowly this function varies throughout field space, it is 
bounded from above by \il{\Gamma \leq \frac{1}{2}} and thus
never can admit stable tracking solutions.  

On the other hand, allowing for \il{mR>0} sufficiently large 
such that \il{e^{-2\pi mR}\ll 1}, we can approximate 
\beq\label{eq:clockworkGamma}
\Gamma \approx 1 - \frac{1}{2}\sech^2\!\!\left(\frac{\phi}{2f}\right) < 1 \ .
\eeq

Although $\Gamma$ is always less than unity, for field values larger than \il{\phi/f\gtrsim 3} we 
can achieve \il{\Gamma > \frac{9}{10}} and thus find stable tracking solutions.
This is easy to accomplish for \il{\phi \sim f_{\text{eff}}} if $mR$ is moderately large. 
Additionally, we must check that $\Gamma$ is slowly varying over a Hubble time:
\beq
\abs{\frac{1}{\Gamma}\frac{d\Gamma}{dN_e}} 
\approx \abs{\sech\!\left(\frac{\phi}{f}\right)}\tanh^2\!\!\left(\frac{\phi}{2f}\right) \ll 1 \ ,
\eeq 
where $N_e$ is the number of efolds and we used Eq.~\eqref{eq:trackingcond}.  
In the field range where Eq.~\eqref{eq:clockworkGamma}
is viable, the above condition is easily satisfied as well, and we can therefore
\emph{always identify a tracking region of the CCW axion potential} for \il{e^{-2\pi mR}\ll 1}.

Indeed, the above analysis confirms the findings of our numerical simulations in Sec.~\ref{sec:AnExplicitModel},
including the fact that the axion equation of state always appears radiation-like
during the tracking period.  
Using Eq.~\eqref{eq:trackingw}, we find 
\beq
\wphi \approx w + \left(1+w\right)\csch^2\!\left(\frac{\phi}{2f}\right) \ ,
\eeq
which in the proper field range matches the background 
\il{\wphi \approx w} to an excellent approximation.

\FloatBarrier
\section{Boltzmann Equations for \texorpdfstring{$B-L$}{B-L} at High Temperature\label{sec:BoltzmannEquations}}


In this appendix, we derive the effective chemical potential $\muBmL$
used in Eq.~\eqref{eq:muBmL}, taking into account the 
details of sphaleron transitions in the Boltzmann evolution.  
To begin, let us consider a species $X$ which is in kinetic equilibrium 
at temperature $T$.  Assuming some chemical potential $\mu_X$,
the asymmetry in number density between particles and antiparticles 
is described by either Fermi-Dirac $(+)$ or Bose-Einstein $(-)$ statistics as
\begin{align}
    n_X = g_X \int \frac{d^3 \vec p}{(2\pi)^3} &\left[\frac{1}{\exp\left[\left(E_X - \mu_X\right)/T\right] \pm 1}\right. \nn \\
    &\hspace{2mm}-\left.\frac{1}{\exp\left[\left(E_X + \mu_X\right)/T\right] \pm 1}\right] \ , 
\end{align}
where $g_X$ is the number of degrees of freedom for the species
and \il{E_{X}=\sqrt{p^2 + m_X^2}} is the energy. 
At high temperature \il{T\gg (\mu_X, m_X\!)}, this is well-approximated by
\beq
    n_X \approx\left\{
        \begin{array}{ll} 
        g_X \mu_X T^2/6 & \ \ \text{for fermions} \\
        g_X \mu_X T^2/3 & \ \ \text{for bosons}
        \end{array}\right. \ ,
\eeq
such that a proportionality exists between the chemical potential 
and the number density for the species.

Let us now consider that this species is involved in some chemical process
$A$, according to
\beq
A: \ \ X + i +\cdots \ \longleftrightarrow \ j + \cdots \ .
\eeq
Naturally, if the reaction is sufficiently rapid and 
it reaches chemical equilibrium, then the associated chemical potentials 
satisfy algebraic relations
\beq\label{che_eq}
d_{A,X}\, \mu_X + d_{A, i}\, \mu_i +  d_{A, j}\, \mu_j+\cdots = 0 \ ,
\eeq
where $d_{A,X}$ ($d_{A,i}$) denotes the multiplicity of $X$ \!($i$)
and the signs determine the direction of the reaction.
In the case of a spatially homogeneous and spontaneous violation of $CPT$ symmetry, 
as studied in this paper, these relations are \emph{sourced} by an effective 
chemical potential $\mu_A$.  That is, we instead have the relations
\beq
d_{A,X} \mu_X + d_{A,i} \mu_i +d_{A,j} \mu_j+ \cdots + \mu_A = 0 \ . 
\eeq
The corresponding out-of-equilibrium evolution for the number density $n_X$ is
given by the Boltzmann equation
\begin{align}\label{che_boltzmann}
    \dot n_X &+ 3 H n_X \\
    &=-\sum_A d_{A,X}\gamma_A\left(d_{A,X}\,\frac{\mu_X}{T}+d_{A,i}\,\frac{\mu_i}{T}+\cdots+\frac{\mu_A}{T}\right) \ . \nn
\end{align}
where $\gamma_A$ is the thermally averaged interaction rate density for 
the process $A$ normalized by $T^3$, and the sum is over all the chemical 
processes involving $X$.  We can solve the coupled Boltzmann equations 
with some set of sources $\{\mu_A\}$ and obtain any of the number densities
or chemical potentials in the process, \eg, the lepton- and baryon-number 
density $n_L$ and $n_B$.
 
Considering that all processes preserve the gauge symmetry 
\il{\SUIII_c\times \SUII_W \times \UI_Y} during baryogenesis, the chemical potentials for the
gauge bosons all vanish, and we can impose other additional constraints.  
In particular, the expectation value for the hypercharge \il{\langle Y\rangle}
over the chemical potentials should vanish:
\beq\label{eq:U1Y}
 \sum_i (\mu_{q_i} + 2\mu_{u_i} - \mu_{d_i} - \mu_{\ell_i} -\mu_{e_i})+ 2\muH = 0 \ ,
\eeq
where, respectively, $i$ is a flavor index, $q$ and $\ell$ are left-handed quark 
and lepton doublets, $u$ and $d$ are right-handed up and down quarks, 
$e$ is a right-handed electron, and $\Higgs$ is the Higgs boson.

Under the above constraint, we can show that the quark number densities evolve according to
\begin{align}\label{eq:full_boltz1}
    \dot n_{q_i}+ 3 H n_{q_i} = 
        &-\frac{\gamma_{\lambda_{u_i}}}{T}\left(\mu_{q_i} -\mu_{u_i} +\muH\right) \nn \\
        &-\frac{\gamma_{\lambda_{d_i}}}{T}\left(\mu_{q_i} -\mu_{d_i} -\muH\right)  \nn \\
        &-2\frac{\gammass}{T} \sum_j \left(2\mu_{q_j}-\mu_{u_j} -\mu_{d_j}\right)  \nn \\
        &-3\frac{\gammaws}{T}\Big[\sum_j \left(3\mu_{q_j} + \mu_{\ell_j}\right) +\muws\Big] 
\end{align}
and 
\begin{align}
    \dot n_{u_i}+ 3 H n_{u_i} = 
        &\phantom{+}\frac{\gamma_{\lambda_{u_i}}}{T}\left(\mu_{q_i} -\mu_{u_i}  +\muH\right)   \nn \\ 
        &+\frac{\gammass}{T}  \sum_j \left(2\mu_{q_j}-\mu_{u_j} -\mu_{d_j}\right)  \nn \\
    \dot n_{d_i}+ 3 H n_{d_i} =  
        &\phantom{+}\frac{\gamma_{\lambda_{d_i}}}{T}\left(\mu_{q_i} -\mu_{d_i} -\muH\right) \nn \\
        &+\frac{\gammass}{T}  \sum_j \left(2\mu_{q_j}-\mu_{u_j} -\mu_{d_j}\right)   \ ,
\end{align}
while the lepton number densities evolve as
\begin{align}\label{eq:full_boltz2}
    \dot n_{\ell_i}+ 3 H n_{\ell_i} = 
        &-\frac{\gamma_{\lambda_{e_i}}}{T}\left(\mu_{\ell_i} -\mu_{e_i} -\muH \right) \nn \\
        &-\frac{\gammaws}{T} \Big[ \sum_j \left(3\mu_{q_j} + \mu_{\ell_j}\right) +\muws \Big] \nn\\
        &-\sum_j\frac{{\gammaLV}_{ij}}{T}\left(\mu_{\ell_i} +\mu_{\ell_j} + 2\muH\right) \nn\\
    \dot n_{e_i}+ 3 H n_{e_i} = &\phantom{+}\frac{\gamma_{\lambda_{e_i}}}{T}\left(\mu_{\ell_i} -\mu_{e_i} -\muH \right) \ . 
\end{align}
In the above, the rate densities 
$\gamma_{\lambda_{u_i}}$, $\gamma_{\lambda_{d_i}}$, and $\gamma_{\lambda_{e_i}}$
correspond to Yukawa interactions in the SM, while the other rate densities $\gammass$ and $\gammaws$
correspond to strong and weak sphalerons.  The source of $\ilBmL$ violation in 
this paper is the Weinberg operator in Eq.~\eqref{eq:wein_op}, 
for which we denote the rate density as ${\gammaLV}_{ij}$. 
The one remaining unspecified quantity $\muws$ is related to the spontaneous breaking of the $CPT$ 
symmetry through Eq.~\eqref{eq:CSterm}.  As the axion field rolls down its potential, 
it induces this effective chemical potential for the weak sphalerons:
\beq
\muws = \partial_0\theta \ .
\eeq

Adding the various contributions from the Boltzmann equations above, we can determine 
the number-density evolution for baryons $n_B$ and leptons $n_L$ as  
\begin{align}\label{eq:boltz_nb_nl}
    \dot n_B + 3 H n_B =
        &-3\frac{\gammaws}{T}\Big[\sum_i (3\mu_{q_i}+\mu_{\ell_i})+\muws\Big] \nn \\
    \dot n_L + 3 H n_L =
        &-3\frac{\gammaws}{T}\Big[\sum_i (3\mu_{q_i}+\mu_{\ell_i})+\muws\Big] \nn \\
        &-\sum_{ij}\frac{{\gammaLV}_{ij}}{T}(\mu_{\ell_i}+\mu_{\ell_j} + 2\muH) \ .
\end{align}
It is instructive to comment on the limit where the weak sphaleron 
rate is negligibly small.  Taking \il{\gammaws\rightarrow 0} in 
these equations, we find that the evolution of $n_B$ 
becomes trivial and that \il{n_B=0} if the initial baryon number is zero.
In this limit, the equation for lepton number also loses source terms, implying 
$n_L$ is also vanishing~\cite{Shi:2015zwa}.

With the hypercharge constraint from Eq.~\eqref{eq:U1Y}, and vanishing 
initial conditions \il{\{\mu_i =0\}}, we can in principle solve 
the coupled Boltzmann equations numerically.  However, we can also
simplify them through some physical considerations. 
Let us assume that the Yukawa interactions for $N_f$ generations of fermions are in
equilibrium, in addition to all gauge interactions and the strong and weak sphalerons.
However, we shall ignore the Yukawa interactions of the remaining \il{3-N_f} generations 
during baryogenesis.  In such a case, baryon and lepton number are mostly generated by 
sphaleron processes in conjunction with axion dynamics, which leads approximately to
the flavor-universal contributions 
\beq
n_{B_i} \simeq \frac{1}{3} n_B \qquad 
n_{L_i} \simeq \frac{1}{3} n_L \ .
\eeq
Furthermore, the interactions that violate $\ilBmL$ are not flavor-diagonal.
Instead, they are flavor-democratic, such that the off-diagonal 
components are determined by the PMNS neutrino mixing matrix. 
We can therefore simplify the $\ilBmL$ rate density to
\beq
{\gammaLV}_{ij} \simeq \gammaLV \ ,
\eeq
for all lepton flavors, where we defined $\gammaLV$ in Eq.~\eqref{eq:gammaLVdef}.

Taking these simplifications into account, we can compute 
the necessary chemical potentials. In particular, for the Higgs we find
\beq
\muH = \frac{(9+N_f)n_L - 9 n_B}{2(3+ 5N_f)T^2} \ ,
\eeq
while for the $N_f$ generations of quarks and leptons with Yukawa
interactions in equilibrium we have
\begin{align}\label{eq:chem_matter_Nf}
\mu_{u_i} &= \frac{(9+N_f)n_L-(6-5N_f)n_B}{2(3+ 5N_f) T^2} \nn \\
\mu_{d_i} &= \frac{(12 + 5N_f)n_B - (9+N_f)n_L}{2(3+ 5N_f) T^2} \nn \\
\mu_{\ell_i} &= \frac{7(1+N_f)n_L-3n_B}{2(3+ 5N_f) T^2} \nn \\
\mu_{e_i} &= \frac{3n_B- (1-3N_f)n_L}{(3+ 5N_f) T^2} \ .
\end{align}
and for the remaining \il{3-N_f} generations:
\begin{align}
    \mu_{u_i} = \mu_{d_i} &= \frac{n_B}{2T^2} \nn \\
    \mu_{\ell_i} &= \frac{n_L}{T^2} \nn \\
    \mu_{e_i} &= 0 \ .
\end{align}
Meanwhile, the chemical potential for the left-handed quark doublets
is independent of $N_f$:
\beq
\mu_{q_i} = \frac{n_B}{2T^2}  \ .
\eeq

The $n_B$ and $n_L$ number densities are related to each other by weak sphaleron processes:
\begin{align}
n_B &= \frac{(18+31N_f-3N_f^2)n_\BmL - 2(3+ 5N_f)\muws T^2}{45 + 73N_f - 3N_f^2} \nn \\
n_L &= \frac{-3(9+14 N_f)n_\BmL - 2(3+ 5N_f)\muws T^2}{45 +73N_f - 3N_f^2} \ .
\end{align}
The evolution of  $n_\BmL$ is determined by the difference 
between the equations in Eq.~\eqref{eq:boltz_nb_nl} and the chemical potentials above:
\beq
    \dot{n}_\BmL + 3Hn_\BmL = -\GammaLV\left(n_\BmL - n_\BmL^{\text{eq}}\right) \ ,
\eeq
where the rate is given by
\beq
\GammaLV = \frac{9(171+65 N_f - 6N_f^2)}{45+73 N_f -3 N_f^2}\frac{\gammaLV}{T^3}
\eeq
and the equilibrium number density is given by
\beq\label{eq:nBmLeqapp}
n_\BmL^{\text{eq}} = -\frac{2(36+65 N_f -6 N_f^2)}{9(171+65 N_f -6 N_f^2)} \muws T^2 \ .
\eeq
The above expression provides us with the coefficient that appears in Eq.~\eqref{eq:muBmL}.
We are now equipped to compute the final number density $n_\BmL$ and therefore
the final baryon asymmetry.  In particular, after the weak sphalerons decouple 
at \il{T\lesssim 100\,\GeV}:
\begin{eqnarray}
n_B =\frac{28}{79} n_\BmL \ .
\end{eqnarray}

\bibliography{references}

\end{document}